%% file: arxiv.tex
\documentclass{article}
\usepackage{arxiv}   
\usepackage{etoolbox}
\AtBeginEnvironment{table}{\small}

\usepackage[T1]{fontenc}
\usepackage{graphicx}
\usepackage{amsmath}
\usepackage{amssymb}
\usepackage{bm}
\usepackage{subcaption}
\usepackage{booktabs}
\usepackage{algorithm}
\usepackage{algpseudocode}

\algnewcommand{\algorithmicbreak}{\textbf{break}}
\algnewcommand{\Break}{\algorithmicbreak}
\algnewcommand{\algorithmiccontinue}{\textbf{continue}}
\algnewcommand{\Continue}{\algorithmiccontinue}

\DeclareMathOperator*{\E}{\mathbb{E}}
\DeclareMathOperator*{\argmin}{arg\,min}
\newcommand{\R}{\mathbb{R}}

\usepackage{xcolor}

\makeatletter
\renewcommand{\paragraph}{%
  \@startsection{paragraph}{4}{\z@}%
                {3.25ex \@plus 1ex \@minus .2ex}%
                {-1em}%
                {\normalfont\normalsize\bfseries}%
}
\makeatother

\newcommand{\ack}[1]{\section*{Acknowledgments}#1}
\newcommand{\roles}[1]{\section*{Author contributions}#1}
\newcommand{\data}[1]{\section*{Data availability}#1}
\newcommand{\widetablesetup}{\footnotesize\setlength{\tabcolsep}{4pt}}

\usepackage[hidelinks]{hyperref}

\begin{document}

\title{Meta-learning accelerates detector design optimization}

\author{%
  Maxim Borisyak\thanks{Author to whom any correspondence should be addressed: \texttt{mborisiak@constructor.university}}$^{1}$
  \quad Nikita Gladin$^{1, 2}$
  \quad Andrey Ustyuzhanin$^{1, 3, 4}$ \\[0.5em]
  $^{1}$Constructor University Bremen \quad $^{2}$University of Bologna \\
  $^{3}$Constructor Labs \quad $^{4}$National University of Singapore
}

\maketitle

\begin{abstract}
\input{abstract}
\end{abstract}


\input{body}

\end{document}

%% file: abstract.tex
The quality of a detector design is ultimately determined by the quality of the inference it enables, that is, by the accuracy with which the quantities of interest are reconstructed from the raw detector response.
For complex detectors, the inference is performed by machine learning models, and the relation between the design and the attainable inference performance is, in general, non-trivial.
In this work, we consider the optimization of the inference performance with respect to the detector design.
The conventional approach prescribes retraining the inference model at every candidate design, thus, treating the evaluations as independent tasks and discarding the shared structure of the optimal inference algorithms at different designs.
We propose the meta-learned objective estimate (MLOE): instead of solving the inference problem anew at every candidate design, a single meta-inference model, conditioned on the design and trained continually along the optimization path, is shared across all of them.
We test MLOE on three families of optimization problems, the last of which comprises two design spaces of the Spectrometer Straw Tracker of the Search for Hidden Particles (SHiP) experiment; under matched budgets of simulation calls, the meta-inference model evaluates a candidate design using fewer simulation calls than the baseline strategies and holds the better rank over the convergence curve in all examined cases.


%% file: body.tex
\section{Introduction}\label{sec:intro}

Practically all detectors have design variables; the exact nature of these variables depends on the detector, the engineering and cost constraints, and the area of application.
Notably, detectors in high-energy physics are famously complex --- they combine a multitude of sensors with the goal of reconstructing the products of a collision of two high-energy particles and, ultimately, of probing the fundamental laws of physics; the design variables include the overall geometry of the detector, the positions and sizes of individual sensors, and the choice of materials.
During the planning of a detector, the task of optimizing its design arises naturally~\cite{dorigo2023toward}.

The purpose of detectors is to enable accurate inference of quantities of interest (QoI) from raw detector responses: the quality of a detector design is, therefore, the quality of the QoI estimator built on the detector's raw measurements.
In many instruments, the QoI is connected to the response by a known physical relation, and the inference is performed by a static, non-trainable estimator.
For example, a time-of-flight detector translates the transit time between two probes into a velocity; the estimator depends on the design, the distance between the probes, but through an expression known in closed form.
In such cases, the quality of a candidate design can then be evaluated by propagating its noise characteristics through this fixed estimator family.

In some cases, the optimal estimator family is not known a priori; however, the detector admits a quality proxy (e.g., a noise level, a signal-to-noise ratio, or a manually designed criterion), computed from the design characteristics and related monotonically to the achievable inference performance; thus, the design can be optimized using the proxy directly.
Examples include the muon shield of the Search for Hidden Particles (SHiP) experiment, optimized with a Gaussian process surrogate over the residual muon flux and the mass of the shield~\cite{baranov2017optimising}, the geometry of its spectrometer tracker, optimized over hand-crafted track reconstruction metrics~\cite{alenkin2019optimization}, and the dual-radiator Cherenkov detector proposed for the Electron-Ion Collider, optimized over particle identification figures of merit~\cite{cisbani2020ai}; gradient-based variants of this scheme differentiate an explicit objective through local generative surrogates of the simulation~\cite{shirobokov2020black}.

In complex detectors, the response is high-dimensional, in high-energy physics, for instance, a sparse collection of signals from individual readout channels, and neither the optimal inference algorithm nor its achievable performance is evident from the design; the inference is, therefore, delegated to trainable machine learning models.
The performance of the resulting system depends on both the design and the inference algorithm, and the quality of a design is, consequently, a function of the inference algorithm that is \emph{optimal} for this design.
In this work, we consider the detector optimization problem in its general form, i.e., we assume that there is no known proxy that tracks the performance of the optimal inference algorithm.

In such cases, the optimization requires estimating, for every candidate design, the performance achievable by a trained inference algorithm.
The estimate is obtained on samples drawn from a simulation, and the simulations involved, whether tracking particles through matter and electromagnetic fields or integrating the kinetics of a biochemical process, are typically computationally heavy.
Moreover, the simulations rarely provide gradients with respect to the design, so the minimization proceeds in a black-box manner and requires a large number of candidate evaluations.
The cost of an optimization run, thus, has two major components: the simulation of training samples and the training of the inference models. The former usually dominates; throughout this work, we, therefore, consider the convergence of optimization procedures with respect to the total number of simulation calls.

Most general-purpose optimization algorithms are formulated for an objective accessible through an evaluation oracle, a procedure that returns the (possibly stochastic) value of a fixed function at any queried point.
A change in the design configuration, in general, shifts both the distribution of detector responses and the optimal mapping from responses to the QoI, and, therefore, warrants retraining of the inference model at every candidate design.
In this work, we argue that conventional approaches to estimating the objective function waste a significant number of training samples, and, thus, computational resources, on retraining the inference model.
We notice that, in practical cases, the optimal inference algorithms at different designs are far from independent: the optimal estimates change smoothly with the design and, for example, models trained on samples drawn under neighboring designs are strongly correlated.
Each retraining, thus, relearns, at full sample cost, information already contained in the models trained for previously visited designs, while the number of simulation calls consumed per evaluation directly limits the number of optimization steps affordable within a fixed budget~\cite{borisyak2020adaptive}.

We propose to exploit the correlation between designs explicitly by casting inference as a meta-learning problem~\cite{finn2017model, sung2018learning, garnelo2018conditional}.
We call this the \emph{meta-learned objective estimate} (MLOE): instead of retraining a separate inference model per design, a single \emph{meta-inference model} is trained continually along the optimization path, thus, improving the sample efficiency of the inference training and, therefore, the convergence of the optimization with respect to the total number of simulation calls.

Our contributions are as follows:
\begin{itemize}
    \item we introduce the meta-learned objective estimate, in which a single meta-inference model replaces multiple independently trained ones and improves the overall sample efficiency of the optimization;
    \item we evaluate MLOE numerically against conventional baseline strategies on three numerical experiments, two of which are realistic simulations.
\end{itemize}
In these experiments, the meta-inference model evaluates a candidate design using fewer simulation calls than the baseline strategies, and holds the better rank over the convergence curve in all of the cases examined.

\section{Method}\label{sec:method}

The problem of detector design optimization can be formally stated as follows:
\begin{gather}
    \min_{\theta} L(\theta); \notag \\
    L(\theta) = \E_{x, y \sim P(x, y \mid \theta)} l\bigl(f^*_\theta(x), y\bigr) + R(\theta);
    \label{eq:loss}
\end{gather}
where $\theta \in \Theta \subset \R^d$ --- the design parameters, $P(x, y \mid \theta)$ --- the joint distribution of the detector response $x$ and the quantity of interest $y$ under the design $\theta$, $f^*_\theta$ --- the optimal reconstruction algorithm for the design $\theta$, $l(\hat y, y)$ --- the reconstruction loss of the estimate $\hat y$, and $R(\theta)$ --- an optional regularization term, e.g., the cost of the design.

The distribution $P(x, y \mid \theta)$ is not available in analytical form; it is accessible only through sampling from a computationally expensive simulation.
In our experiments, the design constraints, when present, are expressed as a penalty term in $R(\theta)$.

Assuming that $f^*_\theta$ belongs to a parametrized family $f_W$ (for example, a neural network), the problem~\eqref{eq:loss} can be rewritten as a nested one:
\begin{equation}
    \min_\theta L(\theta) = \min_{\theta} \left[\min_W \E_{x, y \sim P(x, y \mid \theta)} l\bigl(f_W(x), y\bigr) + R(\theta)\right].
    \label{eq:joint}
\end{equation}
Throughout the paper, we assume that the family $f_W$ is represented by a neural network; however, many other machine learning algorithms can be used instead.

Solving the problem~\eqref{eq:joint} jointly over $(\theta, W)$ is challenging, since the gradient $\nabla_\theta L$ involves the derivative of $P(x, y \mid \theta)$ with respect to the design $\theta$, which is typically not available for Monte Carlo simulations.
Using black-box optimization methods to optimize $\theta$ and $W$ simultaneously is practically infeasible, since even relatively small machine learning models have far too many parameters for black-box methods.
Thus, a conventional approach is to solve the problem~\eqref{eq:joint} in its nested form: the outer (design) optimization is carried out in a black-box manner, while the model training is performed by a gradient or quasi-Newton optimizer.

In practice, training of the inference model is done on a finite sample and the inner optimization problem is solved up to a certain precision.
This precision directly affects the quality of the outer optimization; thus, it is typically required to be small relative to the range of the objective function.
For a general-purpose machine learning model, this naturally requires a large number of $(x, y)$ samples.
The straightforward way to evaluate $L(\theta)$ at a candidate design is to train $f_W(\cdot)$ from scratch on a fresh dataset $x_i, y_i \sim P(x, y \mid \theta)$, which keeps the estimates at different designs independent of one another. Such an approach is common for existing black-box schemes for design optimization~\cite{shirobokov2020black}.
The same protocol is the standard full-fidelity evaluation in hyperparameter optimization and neural architecture search~\cite{snoek2012practical, zoph2017neural, elsken2019neural}, where its cost is the principal motivation for multi-fidelity alternatives~\cite{li2018hyperband, falkner2018bohb}.
For each evaluated design, such a procedure draws the sample size required to train an inference model from scratch, which, for realistic detectors and simulators, is prohibitive~\cite{cisbani2020ai, fanelli2023ai, figalli2026designing}.

Warm-starting the training from the weights obtained at previous iterations is a standard device in searches over configurations~\cite{yosinski2014transferable, chen2016net2net, real2017large, jaderberg2017population}, and, for detector reconstruction specifically, a particle-flow model pretrained on one detector geometry has been reported to reach the accuracy of from-scratch training on a new geometry with an order of magnitude fewer samples of the latter~\cite{mokhtar2025finetuning}.
Such a strategy is, however, not neutral with respect to the loss estimate: warm-started networks have been shown to generalize worse than freshly initialized ones at matched training loss~\cite{ash2020warm, berariu2021study, shin2024dash}, so the value reported at a design depends on the path along which that design was reached.

More explicit transfer learning techniques regularize the network parameters toward the weights from the previous iterations~\cite{kirkpatrick2017overcoming, zenke2017continual, li2018explicit} or share a part of the weights~\cite{pham2018efficient, bender2018understanding}.
Such techniques, while effective in reducing the number of training samples, risk biasing the loss estimate and, thus, the overall design optimization: transfer from an insufficiently related source can degrade, rather than improve, the target performance~\cite{wang2019characterizing}, and, in one-shot neural architecture search, weight sharing has been shown to distort the ranking of candidate architectures to the point where the search performs comparably to random sampling~\cite{yu2020evaluating}.
Any significant bias also violates an assumption common to optimization procedures: the measured objective value becomes dependent on the optimization trajectory.
The only reasonable way to measure or remove this bias is a comparison against the loss of an unconstrained network, which, in our setting, defeats the purpose of the technique.

Over the course of the optimization, the optimizer probes a number of designs $\theta_i$ and collects a response-target dataset for each of them. This is the setting of meta-learning, in which a single model is trained on a family of related tasks rather than on a single one~\cite{hospedales2022meta, finn2017model, sung2018learning, garnelo2018conditional} --- each detector design defines a separate inference task, the distribution $P(x, y \mid \theta_i)$ paired with the loss $l$, and the designs visited along the optimization path form the distribution of tasks.

Meta-learning models have been repeatedly shown to require fewer samples per individual task than models trained on each task separately, both when the knowledge shared across tasks is an initialization~\cite{finn2017model, nichol2018first} and when it is a conditional model evaluated in a single forward pass~\cite{garnelo2018conditional, garnelo2018neural}.
The effect is also supported theoretically: for a family of tasks admitting a common representation, the number of samples required per task to reach a given accuracy decreases as the number of tasks grows, provided the tasks are sufficiently diverse~\cite{baxter2000model, maurer2016benefit, tripuraneni2020theory}.

\begin{figure}
\centering
\begin{subfigure}{0.49\textwidth}
    \centering
    \includegraphics[width=\linewidth]{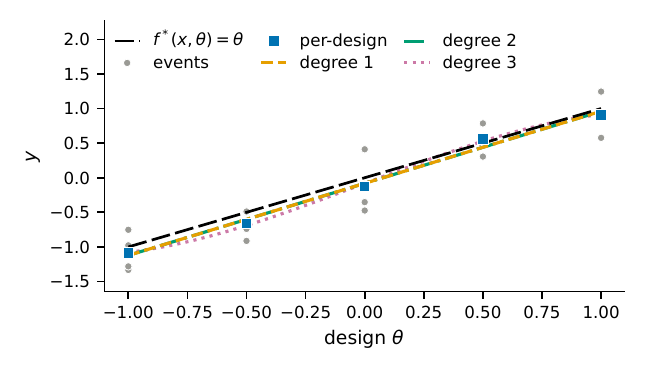}
    \caption{One realization, 5 designs with 4 events each.}
    \label{fig:demo-fits-5}
\end{subfigure}
\hfill
\begin{subfigure}{0.49\textwidth}
    \centering
    \includegraphics[width=\linewidth]{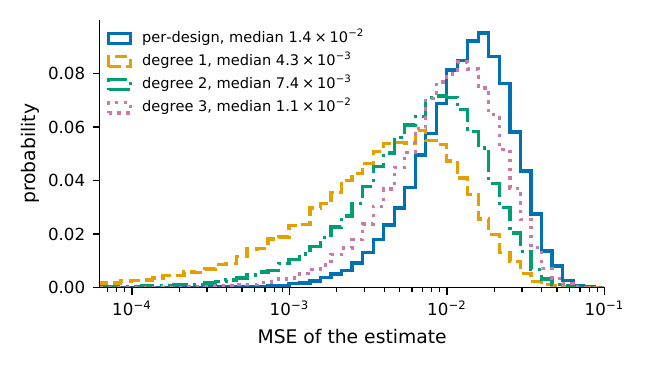}
    \caption{Distribution of the estimation error, 5 designs.}
    \label{fig:demo-mse-5}
\end{subfigure}
\\[1ex]
\begin{subfigure}{0.49\textwidth}
    \centering
    \includegraphics[width=\linewidth]{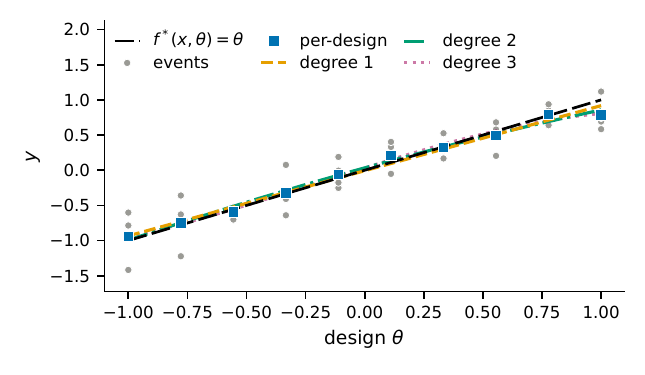}
    \caption{One realization, 10 designs with 4 events each.}
    \label{fig:demo-fits-10}
\end{subfigure}
\hfill
\begin{subfigure}{0.49\textwidth}
    \centering
    \includegraphics[width=\linewidth]{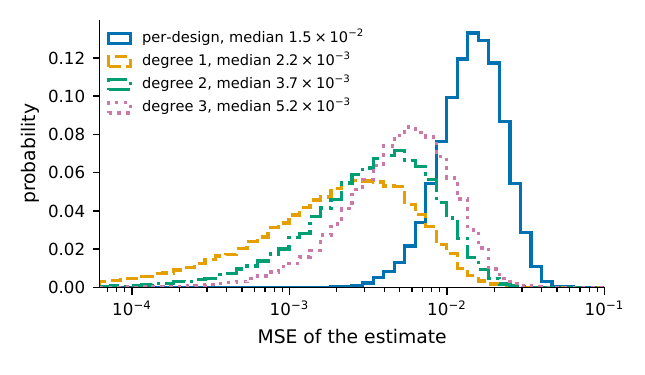}
    \caption{Distribution of the estimation error, 10 designs.}
    \label{fig:demo-mse-10}
\end{subfigure}
\caption{Illustration of the meta-learning effect for an explicitly known design. In this illustration, the event is empty, $x \in \R^0$, and the target is equal to the design itself. Conventional approaches estimate $f$ at each design individually (per-design). The proposed method amounts to a regression in the extended space $f(x, \theta)$, here demonstrated with polynomial models of several degrees: under a suitable inductive bias, the estimate at one design benefits from the events simulated at all the others. The effect persists even when the inference model is more flexible than the ground truth, and becomes more pronounced as the number of designs grows.}
\label{fig:demo-pooling}
\end{figure}

Our case has two particularities. Firstly, the task descriptor, a quantity that uniquely identifies the task within the family, is known exactly --- the design $\theta$ fully specifies the inference problem $P(x, y \mid \theta)$.
In this setting, meta-learning approaches that infer a task embedding from the data reduce to conditioning on the design~\cite{garnelo2018conditional, baldi2016parameterized}.
This suggests a single meta-inference model shared across designs that receives the design parameters as an additional argument:
\begin{equation}
    \min_\theta L(\theta) = \min_{\theta} \min_{W} \left[\E_{x, y \sim P(x, y \mid \theta)} l\bigl(f_W(x, \theta), y\bigr) + R(\theta)\right].
    \label{eq:meta}
\end{equation}
With the design provided explicitly, the meta-learning setting can be interpreted as a regression extended by the design variables; a simple example of this effect is illustrated in figure~\ref{fig:demo-pooling}.

Secondly, the family of tasks is fixed, but the distribution over it is not: the tasks presented to the model are those probed by the optimizer, so the distribution first widens as the search explores the design space and then concentrates as it converges.
To address the distribution shift, we propose to continually retrain the meta-inference model as new tasks become available, i.e., fit the inference model to \emph{all observed data}:
\begin{equation}
    W^* = \argmin_{W} \sum^N_{i = 1}\sum^{m_i}_{j = 1} l\bigl(f_W(x_{i, j}, \theta_i), y_{i, j}\bigr); \label{eq:meta-loss}
\end{equation}
where $N$ is the current iteration of the optimization, $m_i$ the number of samples simulated at the $i$-th iteration.

Note that the procedure~\eqref{eq:meta-loss} resembles transfer learning; however, transfer learning methods usually pull the parameters toward the solution obtained at the previously visited design and generally lead to a residual bias~\cite{ash2020warm, wang2019characterizing, yu2020evaluating}. The proposed procedure instead \emph{retrains} the inference model to be optimal on all visited designs at once; with the design entering as an explicit argument, given sufficient capacity and a sufficient number of samples, the fitted model achieves the per-design optimum at $\theta_i$, so the estimate carries no bias from the designs visited before it.

\paragraph{Structured mini-batch sampling.} We additionally propose a modification that accelerates training of the inference model --- constant weighting of the loss on the accumulated samples, which we refer to as the \emph{context}, against the loss on the current design:
\begin{align}
    W^* &= \argmin_{W} \mathcal{L}; \\
    \begin{split}
        \mathcal{L} &= \mathcal{L}_N + \lambda\,\mathcal{L}_{< N} \\
        &= \bigl(m_N\bigr)^{-1}\sum^{m_N}_{j = 1} l\bigl(f_W(x_{N, j}, \theta_N), y_{N, j}\bigr)
        + \lambda \left(\sum^{N - 1}_{k = 1} m_k\right)^{-1} \sum^{N - 1}_{i = 1} \sum^{m_i}_{j = 1} l\bigl(f_W(x_{i, j}, \theta_i), y_{i, j}\bigr);
    \end{split} \label{eq:weighted}
\end{align}
where $\lambda$ is a hyperparameter that regulates the weight of the context samples ($\lambda = 1$ is used in our experiments). In mini-batch training, the loss~\eqref{eq:weighted} amounts to drawing a fraction $1 / (1 + \lambda)$ of every mini-batch from the dataset of the current design and the remaining $\lambda / (1 + \lambda)$ from the context, or drawing equal-size batches with weights $1$ and $\lambda$.
This procedure departs from the typical meta-learning settings, where the loss across different tasks is typically averaged with equal weights; however, it alleviates several practical issues that arise from the uniform averaging.
The procedure aims at obtaining the optimal network for all observed designs; however, the current design $\theta_N$ is still of the primary interest. Assuming the network was trained until convergence on the previous iterations, the context samples contribute little to the gradient, while the samples of the current design generally do (unless the inference network is already nearly optimal for all designs); under uniform averaging the latter enter with a weight that is roughly $m_N / M$, where $M$ is the total cumulative number of samples, and this weight decreases as the context grows. In contrast, a constant $\lambda$, combined with structured sampling keeps their contribution fixed over the whole run, so that the rate at which the model adapts to a new design does not degrade. Appendix~\ref{app:weighting} provides a more formal derivation of the argument.

\paragraph{Objective estimation with controlled precision.}
The advantage of the proposed method lies in its sample efficiency. For the purposes of investigating the proposed mechanism, we assume that the black-box optimization algorithm requires low-noise estimates of the objective function. To measure the sample efficiency isolated from other factors, we use a simple training procedure that gradually acquires training samples until a predefined precision is achieved. This choice is not inherent to MLOE: the objective itself could be made adaptive, in the manner of adaptive divergences~\cite{borisyak2020adaptive}, and the estimates could equally be supplied to a Bayesian optimizer that selects the precision of each evaluation~\cite{picheny2013quantile, kandasamy2017multi}. The details of the procedure are given in appendix~\ref{app:estimation}; algorithm~\ref{alg:meta} shows the complete loop and the estimation procedure.

Additionally, we consider the effects of gradually increasing the training dataset during network training; in particular, several studies have shown a loss of plasticity in neural networks, i.e., their ability to adapt to new data~\cite{ash2020warm, berariu2021study, dohare2024loss, lyle2023understanding, shin2024dash}. Such an effect has been studied in the literature on continual learning. We consider several recommended warm-start training regimes for this problem. Appendix~\ref{app:continual} provides a detailed description.

\begin{algorithm}
\caption{Design optimization with a meta-inference model}
\label{alg:meta}
\begin{algorithmic}[1]
\Require design space $\Theta$, simulator $P(x, y \mid \theta)$, loss $l$, target precision $\varepsilon$, budget $B$, sampling increments $n_\mathrm{train}$ and $n_\mathrm{val}$, $\gamma = n_\mathrm{train} / (n_\mathrm{train} + n_\mathrm{val})$ --- training sample fraction
\Ensure the best design found and its objective estimate
\Function{Fit}{$\theta$, $\mathcal{D}$} \Comment{objective estimation with controlled precision, appendix~\ref{app:estimation}}
    \State $\mathcal{D}_\mathrm{train} \gets \emptyset$
    \State $\mathcal{D}_\mathrm{val} \gets \emptyset$
    \State $g \gets +\infty$
    \While{$|\mathcal{D}| + |\mathcal{D}_\mathrm{train}| < \gamma \, B$ \textbf{and} $g > \varepsilon$}
        \State $\mathcal{D}_\mathrm{train} \gets \mathcal{D}_\mathrm{train} \cup \Call{Sample}{\theta, n_\mathrm{train}}$
        \State $\mathcal{D}_\mathrm{val} \gets \mathcal{D}_\mathrm{val} \cup \Call{Sample}{\theta, n_\mathrm{val}}$
        \State $W \gets \Call{Train}{\theta, \mathcal{D}_\mathrm{train}, \mathcal{D}}$ \Comment{equation~\eqref{eq:weighted}}
        \State $L_\mathrm{train} \gets \Call{Evaluate}{W, \mathcal{D}_\mathrm{train}}$ \Comment{loss estimate}
        \State $L_\mathrm{val} \gets \Call{Evaluate}{W, \mathcal{D}_\mathrm{val}}$
        \State $g \gets |L_\mathrm{val} - L_\mathrm{train}|$
    \EndWhile
    \State \Return $g \leq \varepsilon$, $(L_\mathrm{train} + L_\mathrm{val}) / 2$, $\mathcal{D}_\mathrm{train}$
\EndFunction
\Statex
\For{$i = 1, 2, \dots$}
    \State $\theta_i \gets \Call{Propose}{\bigl\{(\theta_j, \hat L_j)\bigr\}_{j < i}}$ \Comment{Bayesian Optimization}
    \State $\mathit{converged}, \hat L_{0, i}, \mathcal{D}_i \gets \Call{Fit}{\theta_i, \bigcup_{j < i} \mathcal{D}_j}$
    \State $\hat L_i \gets \hat L_{0, i} + R(\theta_i)$ \Comment{regularization term of~\eqref{eq:loss}}
    \If{\textbf{not} $\mathit{converged}$} \Comment{budget exhausted}
        \State \Return $\theta_{i^*}, \hat L_{i^*}$; \quad $i^* = \argmin_{j < i} \hat L_j$
    \EndIf
\EndFor
\end{algorithmic}
\end{algorithm}

\section{Related work}\label{sec:related}

\paragraph{Black-box detector optimization.}
Bayesian Optimization with Gaussian process surrogates is a standard tool for tuning detector geometries and sensor placements~\cite{mockus1991bayesian, williams2006gaussian, snoek2015scalable}. In the SHiP experiment, both the muon shield~\cite{baranov2017optimising} and the straw-tube spectrometer~\cite{alenkin2019optimization} were optimized with Gaussian process surrogates over hand-crafted quality metrics. The convergence of such methods depends critically on the choice of the surrogate~\cite{bull2011convergence}, and each objective evaluation hides a full training run of the reconstruction algorithm. Local generative surrogates, L-GSO~\cite{shirobokov2020black}, approximate the simulation itself in the neighborhood of the current design and propagate gradients through the learned generator; mutual-information and diffusion-based surrogates extend this direction~\cite{wozniak2025end, schmidt2025end}. A more complete overview of machine learning for detector design can be found in~\cite{fanelli2022design, dorigo2023toward, aehle2025progress, figalli2026designing}.

\paragraph{Precision-controlled evaluation.}
Several Bayesian Optimization variants can also be used to increase sample efficiency: multi-fidelity methods select the fidelity, and, thus, the cost, of every measurement~\cite{kandasamy2017multi}, quantile-based approaches request evaluations at a tunable precision~\cite{picheny2013quantile}, and freeze-thaw and bandit-based methods terminate unpromising evaluations early~\cite{swersky2014freeze, falkner2018bohb}. Such optimizers could directly couple with our precision-controlled estimates; we use plain Bayesian Optimization with a constant target precision to isolate the effect of MLOE.

\paragraph{Sample-efficient training for black-box optimization.}
Adaptive divergence~\cite{borisyak2020adaptive} reduces the number of simulation calls consumed per evaluation in adversarial optimization by adjusting the capacity of the discriminator to the difficulty of the comparison; our work adopts the same protocol, namely, convergence measured against the number of simulation calls, but targets the inference model rather than the divergence estimator.

\paragraph{Differentiable pipelines.}
The MODE (Machine-Learning Optimized Design of Experiments) collaboration~\cite{dorigo2023toward, aehle2025progress} pursues fully differentiable simulation and reconstruction chains, which permit direct gradient-based design optimization when a faithful differentiable simulator can be constructed. A recent analysis shows that optimizing hardware and inference sequentially is provably suboptimal compared to joint co-design~\cite{dorigo2026codesign}. Our setting is complementary: the simulators remain black-box, and the coupling between design and inference is captured by the meta-inference model rather than by gradients through the simulation.

\paragraph{Meta-learning.}
Meta-learning studies models that solve families of related tasks and adapt to new tasks from few samples, via learned initializations~\cite{finn2017model, finn2018probabilistic, nichol2018first}, learned metrics and relations~\cite{koch2015siamese, sung2018learning}, or conditional architectures~\cite{garnelo2018conditional, garnelo2018neural}. In detector optimization the task description, the design vector, is known exactly, so approaches that infer a task embedding from data reduce to conditioning on the design~\cite{garnelo2018conditional}; in high-energy physics, the same mechanism appears as parameterized networks that receive physics parameters as additional inputs~\cite{baldi2016parameterized}. The meta-inference model is the simplest member of this family, a network receiving the design as an input. More specialized architectures, for example, HyperNetworks~\cite{ha2017hypernetworks, galanti2020modularity, Oswald2020Continual}, have favorable expressivity properties for meta-learning tasks; we choose a plain network for a clean comparison with the baseline strategies, which, generally, do not benefit from such architectures.

A related form of amortization, weight sharing across candidate architectures, is used in one-shot neural architecture search~\cite{pham2018efficient, bender2018understanding, yu2020evaluating}.

\paragraph{Meta-learning for Bayesian optimization.}
A parallel line of work applies meta-learning to the outer optimizer: the results on previous related tasks warm-start the search, through informed surrogates, favorable initial configurations or a learned acquisition function~\cite{volpp2020meta}. MALIBO~\cite{pan2024malibo} learns the utility of a query directly across tasks. Such methods are complementary to ours: they take into account dependencies between distinct optimization problems, treating objectives as related black-box functions, while the meta-inference model uses the common structure of the inference models across the tasks induced by the individual designs; both approaches can be used jointly. For our experiments, we use conventional Bayesian optimization with Gaussian processes.

\section{Experiments}\label{sec:experiments}

We conduct three experiments on detector optimization. In all experiments, the inference model is a permutation-invariant set regressor~\cite{zaheer2017deep}; its architecture is described in appendix~\ref{app:architecture}. The objective is estimated by the procedure described in detail in appendices~\ref{app:estimation}~and~\ref{app:continual} at a fixed target precision, and the outer optimizer is the standard Gaussian Process Bayesian Optimization with an RBF kernel; all strategies are compared under matched budgets of simulation calls.

As discussed in section~\ref{sec:intro}, the proposed method is intended for use with complex, computationally expensive simulations; we therefore assume that the total optimization time is dominated by the sampling of training data and measure the performance of the strategies by the cumulative number of simulation calls. For the purposes of comparison, we use computationally inexpensive simulations and evaluate every strategy multiple times. An estimate of resource consumption under a high-fidelity simulation is given in appendix~\ref{app:timing}.

We compare MLOE to two baseline training strategies: the \emph{independent} strategy starts each optimization iteration from independently drawn network parameters, and in the \emph{continual} strategy the network is initialized with the parameters from the previous iteration. Similarly, \emph{MLOE} starts from the parameters of the previous iteration; in addition, its training set accumulates over the whole optimization trajectory.
Within a single run, the networks of all strategies are initialized from the same PRNG seed. Table~\ref{tab:strategies} summarizes the strategies.

Designs collected during optimization are \emph{verified}: by the analytical solution where one exists, and otherwise by retraining the inference models on an independently sampled dataset of the size of the full budget, which is several times larger than the training dataset of any single design under any strategy. For MLOE, the verification does not include the context. Metrics are computed on the losses reported by the verification procedure.

\begin{table}
\caption{Training strategies. $\theta_i, i = 1, \dots, n$ denote the trajectory of the optimizer in the design space. All strategies use the same architecture, optimizer, and stopping procedure.}
\centering
\begin{tabular}{@{}lll@{}}
\toprule
Strategy & Network initialization & Training dataset \\
\midrule
independent & random initialization & $x, y \mid \theta_n$ \\
continual & $W^*_{n-1}$ & $x, y \mid \theta_n$ \\
MLOE & $W^*_{n-1}$ & mixture of $x, y \mid \theta_n$ and accumulated $x, y \mid \theta_{1..n-1}$ \\
\bottomrule
\end{tabular}
\label{tab:strategies}
\end{table}

The \emph{MLOE} strategy maintains a single meta-inference model for the entire optimization run. At each candidate design, mini-batches are drawn in equal proportion from two separate datasets: the events of the current design and the context, the events of all previously visited designs. It is worth noting that the context is reused only for training; the objective estimate is still computed on the current design's samples, so the estimation procedure is identical for all strategies.

The objective estimation procedure has a choice of warm-start training regime that prescribes how the network should be treated after data addition as a remedy against the loss of plasticity~\cite{dohare2024loss}. Before performing the main experiments, we conduct hyperparameter optimization: each strategy is evaluated under all considered warm-start training regimes in three independent runs. The best-performing warm-start training regime is then selected for each training strategy. The detailed description of the considered warm-start training regimes and the hyperparameter optimization results are available in appendix~\ref{app:continual}. We note that such an exhaustive selection performed only for comparing the strategies; in a practical application, its cost would defeat the purpose of the optimization.

For MLOE, we use a batch with an equal count of samples from the current design and the context, and $\lambda = 1$. The latter balances two loss terms of comparable scale; we chose a relatively low weight, compared to the effective weight of a uniform mini-batch, as a safeguard against the case where the network's capacity is insufficient to express the joint optimal solution for all designs along the optimization trajectory.

In the experiments, we evaluate the average sample consumption per design and the convergence curves. The latter are computed as the minimal loss observed along the optimization trajectory up to a certain cumulative budget. Losses at every visited design are re-evaluated by training a network on independently sampled training datasets of a size matching the full budget. We use average curves for illustrations, and for quantitative comparison we compute the integral ranks of the convergence curves: for all runs that shared a PRNG seed, the rank of each strategy is integrated with respect to the cumulative number of simulation calls, normalized by the total budget. The average integral rank of a strategy is the integral rank averaged across independent runs of that strategy.

In addition to the main experiments, we conduct two studies: one varies the weight of the context samples in the structured mini-batch, the other probes MLOE without the context. The experiments and their results are described in appendices~\ref{app:lambda}~and~\ref{app:ablation}. In appendix~\ref{app:source}, we also provide supplementary plots showing that the better sample efficiency is the dominant source of the MLOE advantage.

\subsection{Probes for linear model}\label{sec:linear}
In this experiment, we consider optimizing probe placement for estimating parameters of a linear model. Let $m$ be the dimensionality of the problem:
\begin{align}
	y_i &\sim \mathcal{N}(w \cdot x_i + b, \sigma^2)\;\text{for}\; i=1, \dots, m + 1;\\
	w & \sim \mathcal{N}(0, I_m);\\
	b & \sim \mathcal{N}(0, 1);\\
	x_i &\in [-1, 1]^m;
\end{align}
where the design $\theta = \{x_i\}^{m + 1}_{i = 1}$ is the positions of probes, $y_i$ --- detector responses. The objective of the detector is to estimate $w$ and $b$; the loss is the mean squared error over the $m + 1$ estimated components, $l = \bigl(\lVert \hat w - w \rVert^2 + (\hat b - b)^2\bigr) / (m + 1)$, so that under the priors above an estimator carrying no information about the event attains a loss of 1.

As the design $\theta = \{x_i\}^{m + 1}_{i = 1}$ is permutation-invariant, we use a Deep Set regressor, which estimates $w$ and $b$. MLOE receives events as $\{(x_i, y_i)\}^{m + 1}_{i = 1}$ while the baseline strategies receive a one-hot-encoded index of the probe instead of $x_i$.

The problem admits an analytical loss, allowing us to verify each design. For a design $\theta$, the smallest loss attainable by any estimator is:
\begin{equation}
    L^*(\theta) = \frac{1}{m + 1} \operatorname{tr}\left[\left(\frac{X^\top X}{\sigma^2} + I_{m + 1}\right)^{-1}\right];
    \label{eq:bayes-risk}
\end{equation}
where $X$ is the matrix of probe positions augmented with a column of ones, $I_{m + 1}$ --- unit-diagonal matrix of size $(m + 1) \times (m + 1)$. Every design visited by the optimizer can, thus, be re-valued precisely independently of which values the strategies report for themselves.

\begin{figure}
    \centering
    \includegraphics[width=\textwidth]{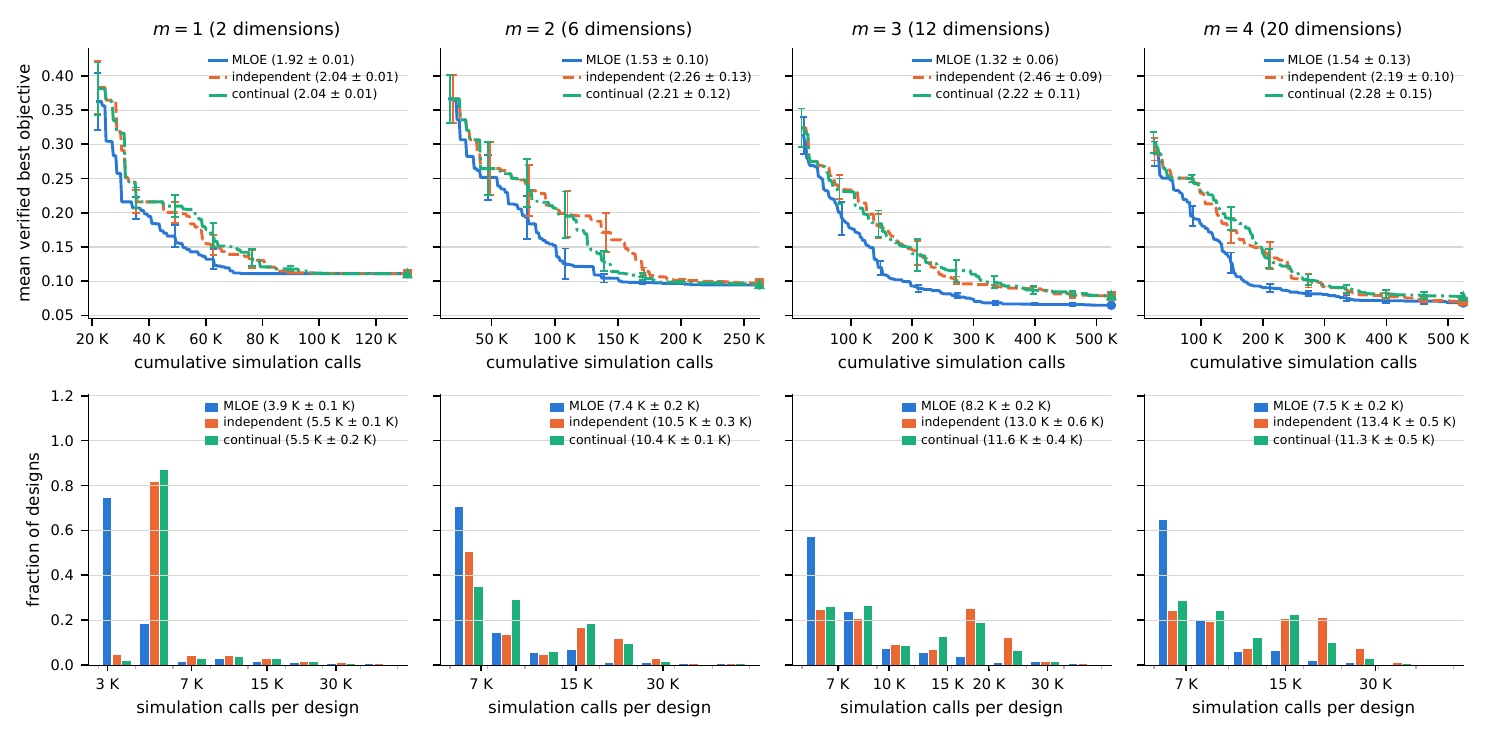}
    \caption{
        Probes for linear model at four dimensionalities $m$, ten seeds each; the design has $m (m + 1)$ dimensions. Upper row: mean verified best objective against the cumulative number of simulation calls; average integral ranks are indicated in parentheses. Lower row: distribution of the simulation calls consumed per design; mean sample consumption per design is indicated in parentheses. The objective is verified by the exact expression for the Bayes risk~\eqref{eq:bayes-risk}.
        For illustration purposes, convergence curves for $m = 1$ and $m = 2$ are cut to $2^{17}$ and $2^{18}$ simulation calls.
    }
    \label{fig:linear-ladder}
\end{figure}

We run the experiment for four dimensionalities, $m = 1, \dots, 4$, at $\sigma = 0.5$, with ten seeds per dimensionality and per strategy. Figure~\ref{fig:linear-ladder} shows the averaged convergence curves. At $m = 1$, the MLOE and independent strategies are statistically close; however, the advantage of MLOE is larger at higher dimensionalities. At $m = 3$ and $4$, MLOE separates from the baselines beyond the standard errors. The continual strategy performs on par with the independent strategy. Additionally, for all $m$, MLOE requires fewer simulation calls than the baseline strategies.

\begin{figure}
    \centering
    \includegraphics[width=\textwidth]{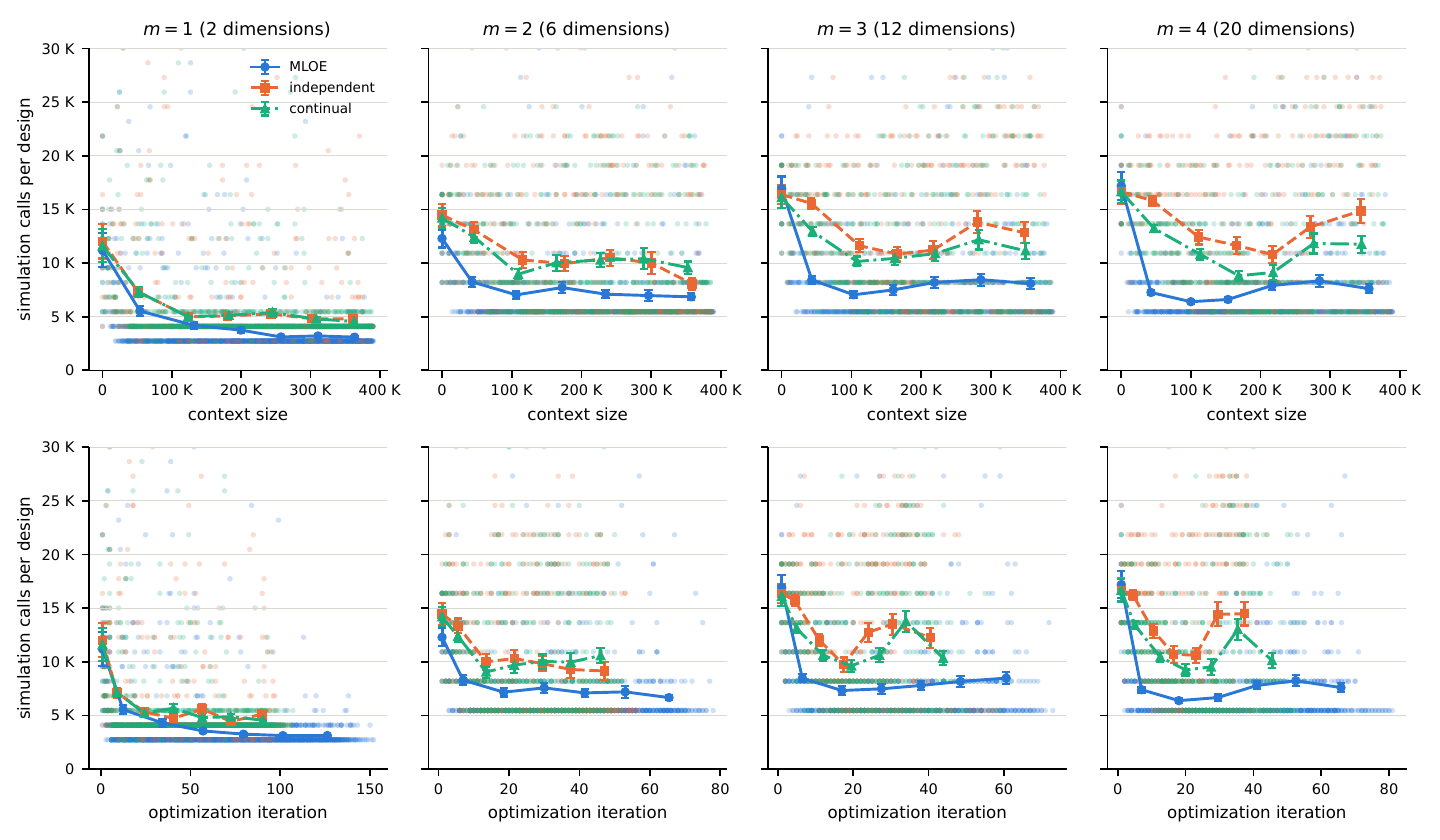}
    \caption{Top: simulation calls per design against the cumulative simulation calls (denoted as context size) on the probes for linear model, one panel per dimensionality $m$. Bottom: the same against optimization iteration. Semi-transparent points: sample consumption on designs; lines: mean $\pm$ standard error in bins of equal count, the first designs are a bin of their own. Ten seeds per dimensionality; 66 of the 8019 designs lie above the frame.}
    \label{fig:context-consumption-linear}
\end{figure}

\begin{table}[ht]
    \caption{Average integral rank on the probes for linear model, by quarter of the simulation budget, over the ten seeds of each dimensionality $m$. The objective is verified by the exact expression for the Bayes risk~\eqref{eq:bayes-risk}. The lowest values are marked in bold, the values within two combined standard errors~---~in italics. The values for $m = 1$ are degenerate after Q1, as all strategies converge on the exact solution within the first quarter.}
    \centering
    \begin{tabular}{@{}llccccc@{}}
\toprule
& Strategy & Q1 & Q2 & Q3 & Q4 & overall \\
\midrule
$m = 1$ & MLOE & $\bm{1.69 \pm 0.05}$ & $\bm{2.00 \pm 0.00}$ & $\bm{2.00 \pm 0.00}$ & $\bm{2.00 \pm 0.00}$ & $\bm{1.92 \pm 0.01}$ \\
 & independent & $2.15 \pm 0.05$ & $\bm{2.00 \pm 0.00}$ & $\bm{2.00 \pm 0.00}$ & $\bm{2.00 \pm 0.00}$ & $2.04 \pm 0.01$ \\
 & continual & $2.16 \pm 0.04$ & $\bm{2.00 \pm 0.00}$ & $\bm{2.00 \pm 0.00}$ & $\bm{2.00 \pm 0.00}$ & $2.04 \pm 0.01$ \\
\midrule
$m = 2$ & MLOE & $\bm{1.35 \pm 0.05}$ & $\bm{1.36 \pm 0.16}$ & $\bm{1.61 \pm 0.18}$ & $\bm{1.79 \pm 0.22}$ & $\bm{1.53 \pm 0.10}$ \\
 & independent & $2.42 \pm 0.04$ & $2.44 \pm 0.17$ & $2.25 \pm 0.24$ & \textit{1.95 $\pm$ 0.27} & $2.26 \pm 0.13$ \\
 & continual & $2.23 \pm 0.04$ & $2.20 \pm 0.16$ & \textit{2.14 $\pm$ 0.22} & \textit{2.26 $\pm$ 0.25} & $2.21 \pm 0.12$ \\
\midrule
$m = 3$ & MLOE & $\bm{1.53 \pm 0.09}$ & $\bm{1.21 \pm 0.10}$ & $\bm{1.18 \pm 0.11}$ & $\bm{1.35 \pm 0.14}$ & $\bm{1.32 \pm 0.06}$ \\
 & independent & $2.31 \pm 0.07$ & $2.57 \pm 0.19$ & $2.57 \pm 0.15$ & $2.39 \pm 0.23$ & $2.46 \pm 0.09$ \\
 & continual & $2.17 \pm 0.07$ & $2.22 \pm 0.15$ & $2.25 \pm 0.21$ & $2.27 \pm 0.24$ & $2.22 \pm 0.11$ \\
\midrule
$m = 4$ & MLOE & $\bm{1.53 \pm 0.08}$ & $\bm{1.27 \pm 0.14}$ & $\bm{1.54 \pm 0.20}$ & $\bm{1.81 \pm 0.26}$ & $\bm{1.54 \pm 0.13}$ \\
 & independent & $2.23 \pm 0.06$ & $2.39 \pm 0.13$ & $2.17 \pm 0.21$ & \textit{1.96 $\pm$ 0.19} & $2.19 \pm 0.10$ \\
 & continual & $2.25 \pm 0.05$ & $2.34 \pm 0.16$ & $2.29 \pm 0.25$ & \textit{2.23 $\pm$ 0.27} & $2.28 \pm 0.15$ \\
\bottomrule
\end{tabular}
    \label{tab:rank-linear}
\end{table}

Table~\ref{tab:rank-linear} illustrates a typical pattern: the rank of MLOE initially improves as the method accumulates context; however, the advantage diminishes as the methods converge to the optimum and become indistinguishable. The latter is most apparent at $m = 1$, where all strategies converge within the first quarter of the budget (the exact solution is reachable since it is on the boundary of the design space). Figure~\ref{fig:context-consumption-linear} (top row) shows sample consumption per design as a function of the cumulative sample size, which is also the size of the context for MLOE. The sample consumption of MLOE generally drops with the increasing context size, however, the baseline methods also show the same pattern, which suggests that the consumption is affected by the inference problem. The latter claim is also supported by the correlated shapes of the consumption curves as functions of the optimization iteration (figure~\ref{fig:context-consumption-linear}, bottom row).

\subsection{Enzyme inhibitor assay}\label{sec:enzyme}
In this experiment, we consider experimental design as a detector, where the design is the full set of experimental conditions executed simultaneously.
We select a typical task from biology: determining the mechanism by which a candidate compound inhibits an enzyme.
The simulation follows a two-substrate enzymatic assay with an inhibitor whose concentration is one of the design parameters
\begin{equation}
    A + B \xrightarrow{\;E\;} C + D;
\end{equation}
the two substrates $A$ and $B$ are turned over into the products $C$ and $D$ by the enzyme $E$, in $1{:}1{:}1{:}1$ stoichiometry.
The kinetic constants and their priors are modelled after the hexokinase family. The inhibitor binds the free enzyme with strength $k_1$, in competition with the product already inhibiting the site of the varied substrate, and the enzyme-substrate complex with strength $k_2$, which caps the turnover. Let $[A]$ and $[B]$ be the concentrations of the two substrates, $[E]$ that of the enzyme and $[I]$ that of the inhibitor; both products have the same concentration $[P]$ and inhibit their respective sites through $K_{i, A}$ and $K_{i, B}$. Let $k_{\mathrm{cat}}$ be the specific reaction rate, and $K_A$, $K_B$ the Michaelis constants of the two substrates, then:
\begin{align}
    K^{\mathrm{app}}_A &= K_A \bigl(1 + [P] / K_{i, A}\bigr); && \text{(product inhibition)} \label{eq:kapp-a} \\
    K^{\mathrm{app}}_B &= K_B \bigl(1 + [P] / K_{i, B} + k_1 [I]\bigr); && \text{(competitive)} \label{eq:competitive} \\
    v &= \frac{k_{\mathrm{cat}} [E] [A] [B]}{\bigl([A] + K^{\mathrm{app}}_A\bigr) \bigl([B] + K^{\mathrm{app}}_B\bigr) \bigl(1 + k_2 [I]\bigr)}. && \text{(uncompetitive)} \label{eq:uncompetitive}
\end{align}
Both mechanisms are present in every compound to a varying degree, and the inhibition class is set by which of $k_1$ and $k_2$ dominates. In this task, we simulate two extreme cases: $k_1 \gg k_2$ and $k_2 \gg k_1$. Every kinetic constant, except for $k_1$ and $k_2$, follows the van~'t~Hoff relation in temperature.

\begin{figure}
    \centering
    \includegraphics[width=\textwidth]{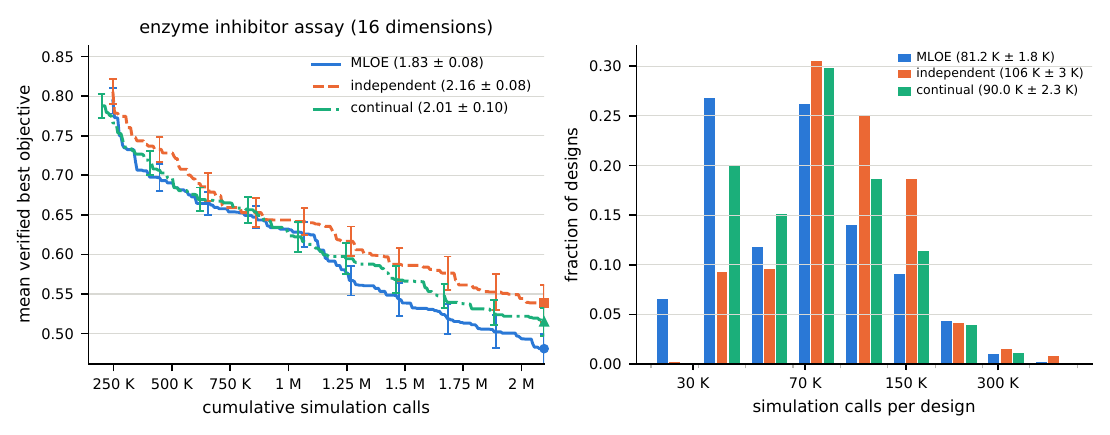}
    \caption{Enzyme inhibitor assay: 16 dimensions, twenty seeds at a budget of $2^{21}$ simulation calls. Left: mean verified best objective against the cumulative number of simulation calls; average integral ranks are indicated in parentheses. Right: distribution of the simulation calls consumed per design; mean sample consumption per design is indicated in parentheses.}
    \label{fig:enzyme-convergence}
\end{figure}

The design is a batch of four experiments, each defined by the enzyme stock dilution, the initial concentrations of $B$ and $I$, and the temperature; $A$ is held fixed and saturating. Typical measurement protocols prescribe four experiments for estimating the type of inhibition, a full set of low and high inhibitor concentrations combined with concentrations of $B$ below and above its Michaelis constant.

The inference network is tasked with classifying the inhibition type from 32 evenly spaced noisy measurements of $[A]$ per experiment.
The inference model is a permutation-invariant set regressor over experimental measurements~\cite{zaheer2017deep}. The experiment is repeated for 20 random seeds.

\begin{figure}
    \centering
    \includegraphics[width=0.55\textwidth]{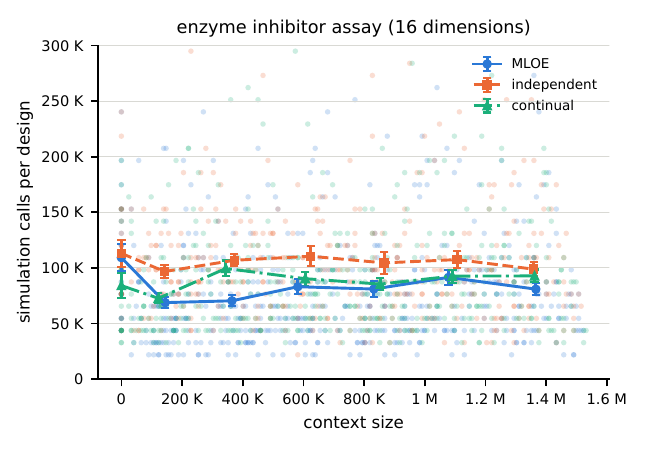}
    \caption{Simulation calls per design against the cumulative simulation calls (denoted as context size) on the enzyme inhibitor assay. Semi-transparent points: sample consumption on designs; lines: mean $\pm$ standard error in bins of equal count, the first designs are a bin of their own. Twenty seeds; twelve of the 1349 designs lie above the frame.}
    \label{fig:context-consumption-enzyme}
\end{figure}

Figure~\ref{fig:enzyme-convergence}~and~table~\ref{tab:rank-extremes} show the results of the experiment. Within each run, MLOE is on par with or better than the continual strategy, while independent shows the worst convergence overall and, at the same time, the best rank in the first quarter (tied within two standard errors), which suggests that the continual strategy exercises a certain degree of transfer between designs. MLOE achieves higher sample efficiency than either of the baseline strategies; however, the advantage is less pronounced than in other experiments. Figure~\ref{fig:context-consumption-enzyme} shows that MLOE sample consumption decreases initially, however, rises slightly (within two standard errors) in the later stages of optimization, which suggests a rising complexity of the classification task as the optimization approaches the optimal design.

\begin{table}[!ht]
    \caption{Average integral rank on the enzyme inhibitor assay, by quarter of the simulation budget, computed on the verified objective values over twenty seeds. The lowest values are marked in bold, the values within two combined standard errors~---~in italics.}
    \centering
    \begin{tabular}{@{}lccccc@{}}
\toprule
Strategy & Q1 & Q2 & Q3 & Q4 & overall \\
\midrule
MLOE & \textit{2.08 $\pm$ 0.09} & $\bm{1.78 \pm 0.11}$ & $\bm{1.80 \pm 0.15}$ & $\bm{1.71 \pm 0.17}$ & $\bm{1.83 \pm 0.08}$ \\
independent & $\bm{1.84 \pm 0.09}$ & \textit{2.07 $\pm$ 0.16} & $2.38 \pm 0.12$ & $2.33 \pm 0.16$ & $2.16 \pm 0.08$ \\
continual & \textit{2.08 $\pm$ 0.09} & \textit{2.15 $\pm$ 0.16} & \textit{1.83 $\pm$ 0.15} & \textit{1.97 $\pm$ 0.15} & \textit{2.01 $\pm$ 0.10} \\
\bottomrule
\end{tabular}
    \label{tab:rank-extremes}
\end{table}

\subsection{Straw spectrometer}\label{sec:tracker}

The simulation follows a straw-tube spectrometer modelled after the Spectrometer Straw Tracker (SST) of the SHiP experiment~\cite{ahdida2022ship, alenkin2019optimization}. The SHiP detector is designed for studying feebly interacting particles. Events are decays of heavy neutral leptons that happen upstream of the SST in the decay volume, a 50~m long vessel whose downstream end lies 0.95~m upstream of the first tracking station. The hypothetical particles decay into two charged daughters; the latter enter and interact with the tracker.

\begin{figure}
    \centering
    \includegraphics[width=0.75\textwidth]{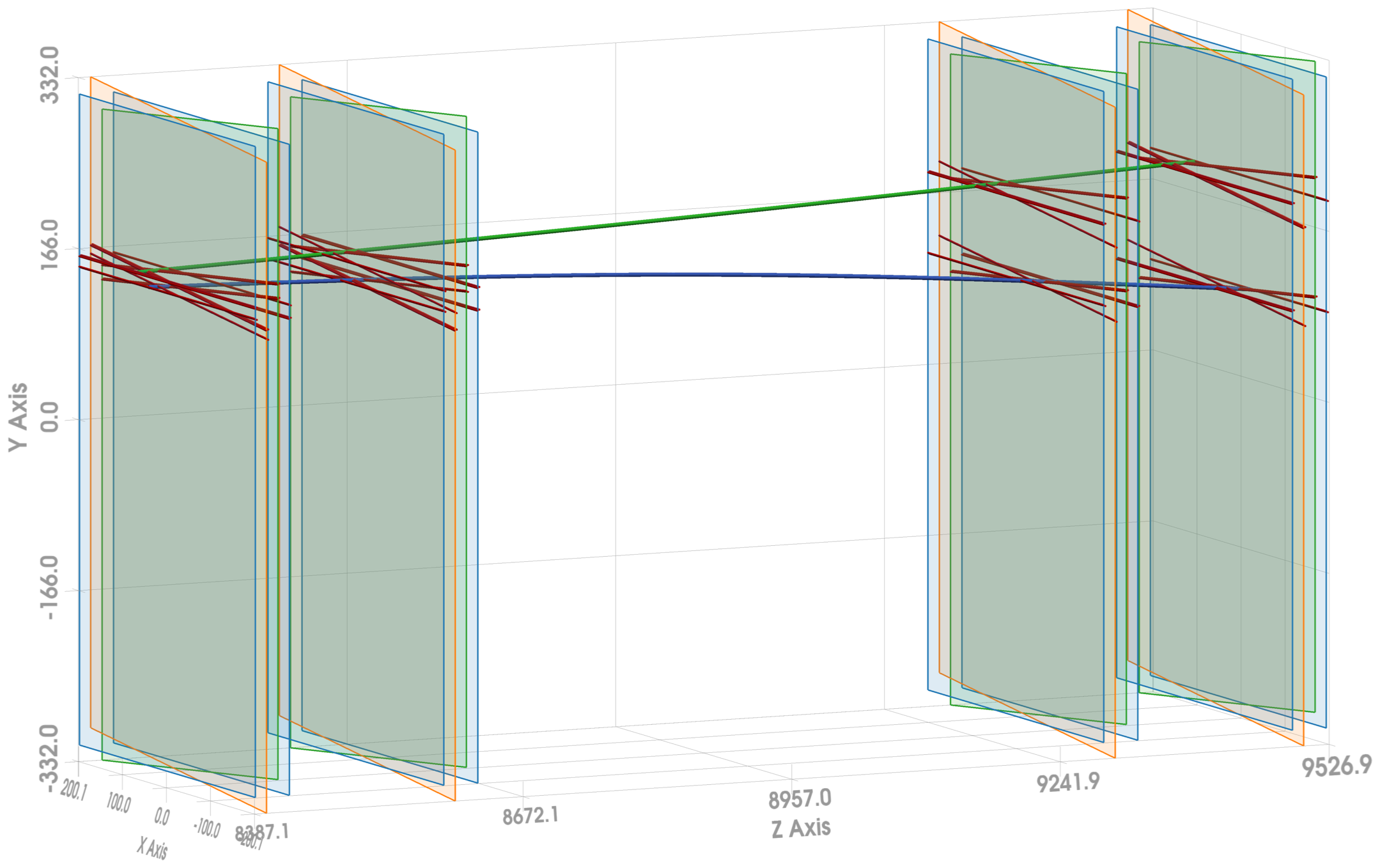}
    \caption{A typical event in the Spectrometer Straw Tracker at the nominal design, viewed from upstream. The layers are drawn as semi-transparent tilted parallelograms, triggered straw tubes are highlighted in red, and the particle trajectories are shown in green and blue. All layers are perpendicular to the $z$ axis.}
    \label{fig:ship-event}
\end{figure}

The SST consists of a magnet and four tracking stations (two upstream and two downstream of the magnet); each station has four stereo views, and each view has two straw layers. The straws are sensitive to charged particles passing through them. The magnet bends the trajectories of the charged particles, and the curvature allows reconstruction of their momenta; the intersection point of the trajectories allows reconstruction of the decay vertex, and the sum of the particle momenta is the momentum of the primary particle. A typical event is shown in figure~\ref{fig:ship-event}.

We use a simplified simulation that reproduces the majority of the physical processes and is validated against FairShip, the Geant4-based simulation framework of the experiment~\cite{ahdida2022ship}. The distribution of incoming particles is not affected by the SST design and was simulated with the aforementioned FairShip Monte Carlo. The response of an event is the set of fired straws, each reported as a discrete address together with a noisy time measurement; responses are capped at 128 hits with the majority of events producing 64 (two charged daughters hitting 32 layers of the detector).

We consider two optimization problems. The first problem, to which we refer as \emph{SST layout}, has $d = 5$ degrees of freedom: the positions of the four stations along the $z$ axis, constrained to either side of the magnet, and one stereo angle, shared by all stations, in $[0, 0.2]$~rad. The remaining geometry (straw pitch, station composition) and the magnetic field strength are fixed. The stereo angle controls the trade-off between resolution in the bending plane and the ability to resolve hits along the $x$ axis; the station positions influence curvature resolution.

The second optimization problem, to which we refer as \emph{SST angle}, is a simplified version of the first one and has a single free design parameter --- the stereo angle, the most influential one. The station positions are fixed at their nominal values, and every other setting is that of the first problem.

\begin{figure}
    \centering
    \includegraphics[width=\textwidth]{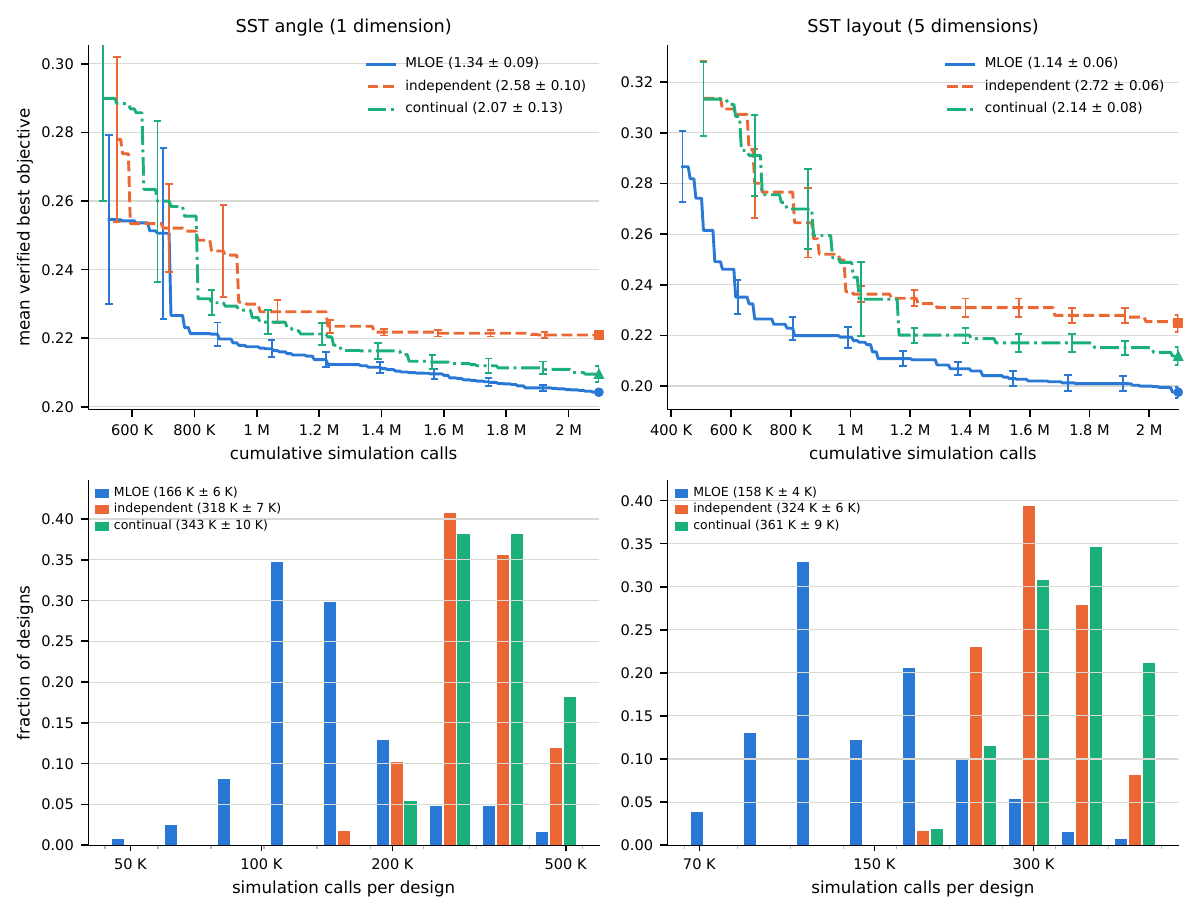}
    \caption{The two design spaces of the SHiP SST. SST angle: the stations are frozen and the stereo angle alone is free, 1 dimension, ten seeds. SST layout: the station positions and the stereo angle are free and overlapping stations are priced, 5 dimensions, ten seeds. Upper row: mean verified best objective against the cumulative number of simulation calls; average integral ranks are indicated in parentheses. Lower row: distribution of the simulation calls consumed per design; mean sample consumption per design is indicated in parentheses.}
    \label{fig:ship-campaigns}
\end{figure}

The reconstruction target is the decay vertex $x_d \in \R^3$, the momenta of the two daughters $p_1, p_2 \in \R^3$, and the momentum of the primary particle $p = p_1 + p_2$. Each component is scaled by the standard deviation computed from a reference sample of the same production. We use the mean squared error (MSE) as the objective function; since the daughters are unordered, the momenta enter through the better of the two assignments. With hats denoting the estimated quantities, the loss reads:
\begin{align*}
    l(\hat{x}_d, \hat{p}_1, \hat{p}_2, x_d, p_1, p_2) &= \frac{1}{2}\|\hat{x}_d - x_d\|^2 \\
    &\quad + \frac{1}{4} \min \left[\|\hat{p}_1 - p_1\|^2 + \|\hat{p}_2 - p_2\|^2, \|\hat{p}_1 - p_2\|^2 + \|\hat{p}_2 - p_1\|^2\right] \\
    &\quad + \frac{1}{4} \left\|(\hat{p}_1 + \hat{p}_2) - (p_1 + p_2)\right\|^2
\end{align*}
For the first task, we additionally introduce a penalty on intersecting stations, proportional to the total length of the intersection. The penalty coefficient is set so that a complete intersection costs approximately as much as the expected best reconstruction error observed on the task.

For the design-independent methods, hits are represented as the response time of the straw and the normalized indices of the station, the view within the station, the layer within the view, and the straw tube within the layer. For the meta-inference model, the design enters the features themselves, which makes the indices redundant~---~we represent hits as the response time of the straw, the normalized $z$ coordinate of its layer, and the left and right $y$ coordinates of the straw tube, i.e., every hit carries the design parameters relevant to the triggered straw tube.

Table~\ref{tab:rank-ship} lists the average integral ranks, and figure~\ref{fig:ship-campaigns} shows the averaged convergence curves. On both tasks, MLOE consumes fewer samples per design, about $160{,}000$ simulation calls against $320{,}000$ and $350{,}000$ for the baseline strategies, which results in faster convergence with respect to the cumulative number of simulation calls. The advantage is more pronounced on the SST layout task, where MLOE reaches good designs early; the gap between the convergence curves diminishes as the rate of improvement slows and all strategies reach the optimal basin.

\begin{figure}
    \centering
    \includegraphics[width=\textwidth]{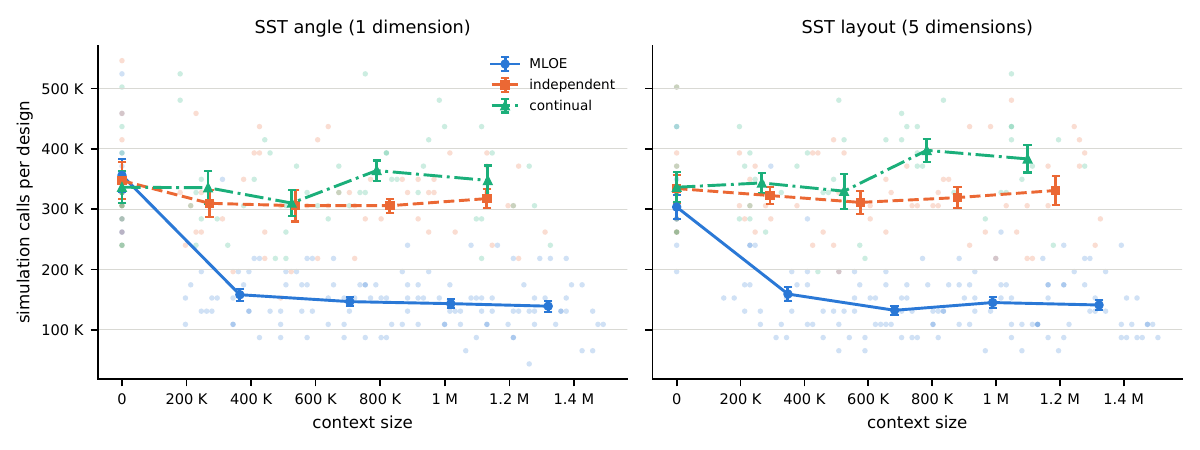}
    \caption{Simulation calls per design against the cumulative simulation calls (denoted as context size). Semi-transparent points: sample consumption on designs; lines: mean $\pm$ standard error in bins of equal count, the first designs are a bin of their own. Ten seeds per strategy per task.}
    \label{fig:context-consumption}
\end{figure}

Figure~\ref{fig:context-consumption} shows per-design sample consumption as a function of cumulative simulation calls, with the latter also being the context size for MLOE. The graphs show that MLOE quickly reduces sample consumption as the context size grows, while the consumption of the baseline strategies stay mostly independent. The continual strategy shows signs of increasing consumption as optimization progressed~---~a potential loss of plasticity related to the phenomenon described in the appendix~\ref{app:continual}.

\begin{table}[!ht]
\caption{Average integral rank on the two design spaces of the SHiP SST, by quarter of the simulation budget, computed on the verified objective values over ten seeds per task. The lowest values are marked in bold, the values within two combined standard errors~---~in italics.}
\centering
\begin{tabular}{@{}llccccc@{}}
\toprule
& Strategy & Q1 & Q2 & Q3 & Q4 & overall \\
\midrule
SST angle & MLOE & $\bm{1.66 \pm 0.18}$ & $\bm{1.17 \pm 0.09}$ & $\bm{1.26 \pm 0.10}$ & $\bm{1.26 \pm 0.13}$ & $\bm{1.34 \pm 0.09}$ \\
 & independent & $2.23 \pm 0.11$ & $2.50 \pm 0.15$ & $2.80 \pm 0.10$ & $2.80 \pm 0.13$ & $2.58 \pm 0.10$ \\
 & continual & \textit{2.11 $\pm$ 0.21} & $2.33 \pm 0.12$ & $1.95 \pm 0.14$ & $1.94 \pm 0.22$ & $2.07 \pm 0.13$ \\
\midrule
SST layout & MLOE & $\bm{1.08 \pm 0.08}$ & $\bm{1.19 \pm 0.10}$ & $\bm{1.15 \pm 0.11}$ & $\bm{1.13 \pm 0.10}$ & $\bm{1.14 \pm 0.06}$ \\
 & independent & $2.49 \pm 0.04$ & $2.62 \pm 0.07$ & $2.90 \pm 0.10$ & $2.87 \pm 0.10$ & $2.72 \pm 0.06$ \\
 & continual & $2.42 \pm 0.11$ & $2.19 \pm 0.14$ & $1.95 \pm 0.16$ & $2.00 \pm 0.15$ & $2.14 \pm 0.08$ \\
\bottomrule
\end{tabular}
\label{tab:rank-ship}
\end{table}

\section{Conclusion}\label{sec:conclusion}

We consider the problem of detector design optimization in which the detector is represented by a stochastic, computationally expensive simulation. For complex detectors, the inference of the quantities of interest often lacks an analytical solution, and machine learning methods are employed. The latter require large datasets for training and evaluation, which leads to a high consumption of computational resources.

Conventional approaches prescribe training an inference model for each evaluated design. We argue that such an approach is sample-inefficient and propose the meta-learned objective estimate that uses a well-known meta-learning effect to reduce the number of samples required for training. The proposed estimate employs a shared network that receives the design explicitly and is trained on the data collected along the whole optimization path. Such an architecture allows the network to learn the shared structure of the optimal inference solutions and reduces the number of samples required for adjusting to a new design.

We evaluated the proposed estimate against two conventional baseline strategies, training the inference model from scratch at every design and continuing from the network of the previous design, on three optimization problems: the placement of probes for a linear model, the design of an assay for classifying the inhibition mechanism of an enzymatic reaction, and the design of a straw tracker modelled after the SST of the SHiP experiment at CERN. In all experiments, MLOE consumes fewer samples per design than the baseline strategies and, consequently, converges faster.

\ack{This research / project is supported by the Ministry of Education, Singapore, under its funding for the Research Centre of Excellence award to the Institute for Functional Intelligent Materials (Project No. EDUNC-33-18-279-V12-IFIM) and by the National Research Foundation, Singapore under its AI Singapore Programme (AISG Award No: AISG3-RP-2022-028).}

\roles{Maxim Borisyak: Conceptualization, Methodology, Software, Validation, Formal analysis, Investigation, Data curation, Visualization, Writing \textendash{} original draft, Writing \textendash{} review \& editing. Nikita Gladin: Software. Andrey Ustyuzhanin: Supervision, Project administration, Writing \textendash{} review \& editing, Funding acquisition.}

\data{The simulation, the optimization pipeline, and the configurations that produce all reported results are openly available at \url{https://github.com/mborisyak/detopt}. The same repository holds the results of the experiments and the Monte Carlo samples from FairShip~\cite{ahdida2022ship}.}

\appendix

\section{Weighting of the context samples}\label{app:weighting}

At the beginning of the $N$-th iteration, the meta-inference model has already been trained on the context, so $\nabla_W \mathcal{L}_{<N} \approx 0$, whereas $\nabla_W \mathcal{L}_N$ might not be small. Assuming that the model has sufficient capacity to achieve minimal loss on all visited designs, uniformly sampling the loss~\eqref{eq:meta-loss} gives the mini-batch gradient estimate $G$:
\begin{align}
    G &= \frac{1}{n} \sum_{i \in \mathcal{B}} g_i; \\
    \mathcal{B} &\sim U^n\bigl\{1, \dots, M_N\bigr\}; \\
    g_i &= \nabla_W l\bigl(f_W(x_i, \theta_i), y_i\bigr);
    \label{eq:dilution}
\end{align}
where $M_m = \sum^m_{k = 1} m_k$ --- the total number of observed samples at the $m$-th iteration.

For the case $m_N \ll M_N$ and a mini-batch of size $n \ll M_N$:
\begin{align}
    \E G &= \frac{m_N}{M_N} \nabla_W \mathcal{L}_N + \frac{M_{N - 1}}{M_N} \nabla_W \mathcal{L}_{<N} \approx \frac{m_N}{M_N} \nabla_W \mathcal{L}_N; \label{eq:uniform-estimate} \\
    \mathrm{Var}(G) &= \frac{1}{n} \mathrm{Var}(g) \approx \frac{1}{n} \mathrm{Var}(g_{<N});
\end{align}
in other words, the variance of the mini-batch estimate remains approximately constant while the magnitude of its mean decreases with the number of samples observed.

Convergence of a stochastic gradient method depends on the noise of the gradient estimate relative to the exact gradient. For a smooth objective and an unbiased gradient estimate of variance $\sigma^2$, the number of stochastic gradient steps required to reach $\E \lVert \nabla f \rVert^2 \leq \delta^2$ grows as $\mathcal{O}\bigl(\sigma^2 \delta^{-4}\bigr)$~\cite{ghadimi2013stochastic}, and no method achieves a better rate~\cite{arjevani2023lower}. In our case, the number of gradient steps required to achieve a fixed precision is $\mathcal{O}\Bigl(\bigl(M_N / m_N\bigr)^4\Bigr)$.

Structured sampling replaces the single mini-batch drawn from the combined dataset by two mini-batches of equal size $n / 2$, one drawn from the current design and one from the context, and applies the weight $\lambda$ to the latter:
\begin{align}
    G_\lambda &= \frac{\hat g_N + \lambda \hat g_{<N}}{1 + \lambda}; \label{eq:structured-estimate} \\
    \hat g_N &= \frac{2}{n} \sum_{i \in \mathcal{B}_N} g_i, & \mathcal{B}_N &\sim U^{n / 2}\bigl\{M_{N - 1} + 1, \dots, M_N\bigr\}; \\
    \hat g_{<N} &= \frac{2}{n} \sum_{i \in \mathcal{B}_{<N}} g_i, & \mathcal{B}_{<N} &\sim U^{n / 2}\bigl\{1, \dots, M_{N - 1}\bigr\};
\end{align}
with the moments:
\begin{align}
    \E G_\lambda &= \frac{\nabla_W \mathcal{L}_N + \lambda \nabla_W \mathcal{L}_{<N}}{1 + \lambda} \approx \frac{1}{1 + \lambda} \nabla_W \mathcal{L}_N; \\
    \mathrm{Var}(G_\lambda) &= \frac{2}{n} \cdot \frac{\mathrm{Var}(g_N) + \lambda^2 \mathrm{Var}(g_{<N})}{(1 + \lambda)^2};
\end{align}
where $\mathrm{Var}(g_N)$ and $\mathrm{Var}(g_{<N})$ are the variances of the per-sample gradients of the two sample sets. Neither moment depends on $M_{N - 1}$: the equal split fixes how many context samples enter a step, and $\lambda$ only rescales their contribution to it. The structured mini-batch, thus, maintains a convergence rate asymptotically independent of the size of the context, and $\lambda$ can be chosen without altering the composition of the batch.

\section{Objective estimation with controlled precision}\label{app:estimation}

We assume that, when the final design is chosen, the inference network is retrained with a large budget, so that the inference errors are as small as the design allows. The optimization therefore has to operate on the loss of the inference model optimal for the current design.

We assume that the optimal loss is the infinite-data limit of finite-sample losses:
\begin{align*}
    L_0(\theta) = \lim_{n \to \infty} L_0(\theta, W^*_n);\\
    W^*_n = \argmin_W \sum^{n}_{i = 1} l\bigl(f_W(x_i), y_i\bigr).
\end{align*}
where $L_0$ denotes the inference term of the objective~\eqref{eq:loss}, so that $L(\theta) = L_0(\theta) + R(\theta)$; the regularization term is a function of the design alone, and the procedure below brackets only $L_0$.

If the training procedure does not have any significant biases, for example, strong regularization, capacity limitations, heavy dropout, etc., one can assume that $L^\mathrm{train}_n \leq L_0(\theta) \leq L^\mathrm{val}_n$~\cite{hastie2009elements, cortes1993learning}; training and validation losses $L^\mathrm{train}_n$ and $L^\mathrm{val}_n$ define a contracting bracket on the limit.

The estimation procedure used in the experiments seeks the limit by gradually growing the datasets and retraining the inference network until the bracket contracts to a predefined $\varepsilon$ precision. The validity of the bracket can be guaranteed only when the training converges; therefore, training and data additions are stopped only when both conditions are satisfied. The limit estimate is:
\begin{align*}
    L_0(\theta) = \frac{1}{2}\bigl(L^\mathrm{train}_n + L^\mathrm{val}_n\bigr);
\end{align*}
We assume that the limit is uniformly distributed within the bracket and report the error of the estimate as:
\begin{align*}
\Delta L_0(\theta) = \frac{1}{\sqrt{12}}\bigl(L^\mathrm{val}_n - L^\mathrm{train}_n\bigr);
\end{align*}
The error is propagated to the Gaussian process.

To detect convergence, we fit epoch losses with Bayesian linear regression; the latter has a centered Gaussian prior with $\sigma$ set to a third of the maximum of observed losses, and is fit with harmonic weights (the $k$-th most recent epoch weighing $1/k$) to enable faster adaptation to the tail of the learning curves while guaranteeing that the posterior concentrates. The linear regression fits only the current training dataset. We note that, when fitting converging monotonic curves~\cite{viering2023shape}, this procedure is conservative: it overestimates the decrease of convex curves and the increase of concave ones since the initial points on the curve induce a diminishing bias, and Bayesian inference with a wide prior prevents the regression from prematurely declaring convergence. Additionally, we introduce an unconditional warmup period of 16 epochs on the linear and enzyme tasks and 2 epochs on the SST tasks.

We also implement preemptive data addition when the bracket is expected to grow significantly with high probability. For the $j$-th training epoch with losses $L^\mathrm{train}(j)$ and $L^\mathrm{val}(j)$, the full procedure reads as follows (in order of priority):
\begin{enumerate}
    \item $j < \mathrm{warmup}$: training;
    \item $P\bigl(L^\mathrm{val}(j + \tau) - L^\mathrm{train}(j + \tau) > \varepsilon\bigr) > \pi$: the bracket is expected to grow beyond the precision~$\rightarrow$~data addition;
    \item $P\bigl(L^\mathrm{train}(j) - L^\mathrm{train}(j + \tau) < \varepsilon\bigr) > \pi$ and $L^\mathrm{val}(j + \tau) - L^\mathrm{train}(j + \tau) > \varepsilon$: training has converged, however, the bracket remains larger than the precision~$\rightarrow$~data addition;
    \item $P\bigl(L^\mathrm{train}(j) - L^\mathrm{train}(j + \tau) < \varepsilon\bigr) > \pi$ and $L^\mathrm{val}(j + \tau) - L^\mathrm{train}(j + \tau) < \varepsilon$: training loss has converged, the bracket is below the precision level~$\rightarrow$~return the estimate;
    \item otherwise continue training;
\end{enumerate}
where $\tau = 16$ is the patience parameter, $\pi = 0.9$ --- confidence level, and the forecasts are made with the linear regression described above. The rule follows the predictive termination criterion of Domhan~et~al.~\cite{domhan2015speeding}, which thresholds the posterior probability that a run will improve on the best result observed so far; as our training is unlimited, the forecasts span $\tau$ epochs.

Table~\ref{tab:growth} summarizes all parameters of the objective estimation procedure used in the experiments.

\begin{table}[!ht]
    \caption{Parameters of the training procedure for each task: the initial training dataset size $n_0$, the increment added at every data addition, the target precision $\varepsilon$, the patience $\tau$, the unconditional warm-up in epochs and the total budget of simulation calls. A quarter of every acquisition is held out for validation.}
    \centering
    \begin{tabular}{@{}lrrrrrr@{}}
        \toprule
        Task & $n_0$ & increment & $\varepsilon$ & $\tau$ & warm-up & budget \\
        \midrule
        probes for linear model, $m = 1$ & 2048 & 1024 & $2 \times 10^{-2}$ & 16 & 16 & $2^{19}$ \\
        probes for linear model, $m = 2, 3, 4$ & 4096 & 2048 & $2 \times 10^{-2}$ & 16 & 16 & $2^{19}$ \\
        enzyme inhibitor assay & 16384 & 8192 & $10^{-2}$ & 16 & 16 & $2^{21}$ \\
        SST angle & 32768 & 16384 & $10^{-2}$ & 16 & 2 & $2^{21}$ \\
        SST layout & 32768 & 16384 & $10^{-2}$ & 16 & 2 & $2^{21}$ \\
        \bottomrule
    \end{tabular}
    \label{tab:growth}
\end{table}

\section{Continual training}\label{app:continual}

Every data addition leaves the network fitted to the previous, smaller dataset. The objective estimation procedure is designed to accurately measure sample consumption in our experiments and reduce simulation calls in practice; the size of the initial training dataset is a priori insufficient to train an accurate inference model. Early data additions might significantly change the loss landscape, leaving the previously fitted network parameters in suboptimal local minima or plateaus, which, in turn, would result in biased objective estimates.

This hypothesis is supported by the literature on continual training, in particular, by Ash and Adams~\cite{ash2020warm}, Berariu~et~al.~\cite{berariu2021study}, and Shin~et~al.~\cite{shin2024dash}, who observe a loss of \emph{plasticity}~\cite{dohare2024loss}: a network continually trained on an expanding dataset generalizes worse than one trained on the same data from a fresh initialization. Lyle~et~al.~\cite{lyle2023understanding} observe a similar phenomenon in Reinforcement Learning settings with shifting rewards. Re-initializing the network at every data addition removes this effect; however, it discards the training progress, which might lead to a waste of computational resources.

In this work, we consider two warm-start training regimes for continual training. We refer to the first as \emph{mixing}, a non-biased version of \emph{L2-SP}~\cite{li2018explicit}. The original \emph{L2-SP} uses an $l_2$ penalty towards the initial parameters, which would bias the train-validation loss bracket; thus, we adopt a non-biased version that directly mixes the network weights $W$ with the initial parameters $W_0$:
\begin{align*}
W' = \rho W + (1 - \rho) W_0.
\end{align*}
We test $\rho = 0.75, 0.9$, and $\rho = 1$ (no mixing, continuing training as is).

The second, \emph{shrink-and-perturb}~\cite{ash2020warm}, shrinks the network parameters and adds an independent re-initialization $\hat{W}$:
\begin{align*}
    W' = \alpha W + \sigma \hat{W}.
\end{align*}
We test $\alpha=0.3$ (one of the recommended values) and $\alpha=0.6$ with $\sigma = 10^{-2}$; $\alpha=1$ coincides with the no-mixing option (except for the noise injection).

To select the best warm-start training regime, we perform an equivalent of hyperparameter selection: we test each warm-start training regime with each training strategy on three independent runs, and \emph{for each training strategy} we select the best warm-start training regime with respect to the integral rank. On the probes for linear model, three runs are performed for each dimensionality, and a common warm-start training regime per strategy is selected across all four dimensionalities.

\begin{figure}[p]
    \centering
    \includegraphics[width=\textwidth]{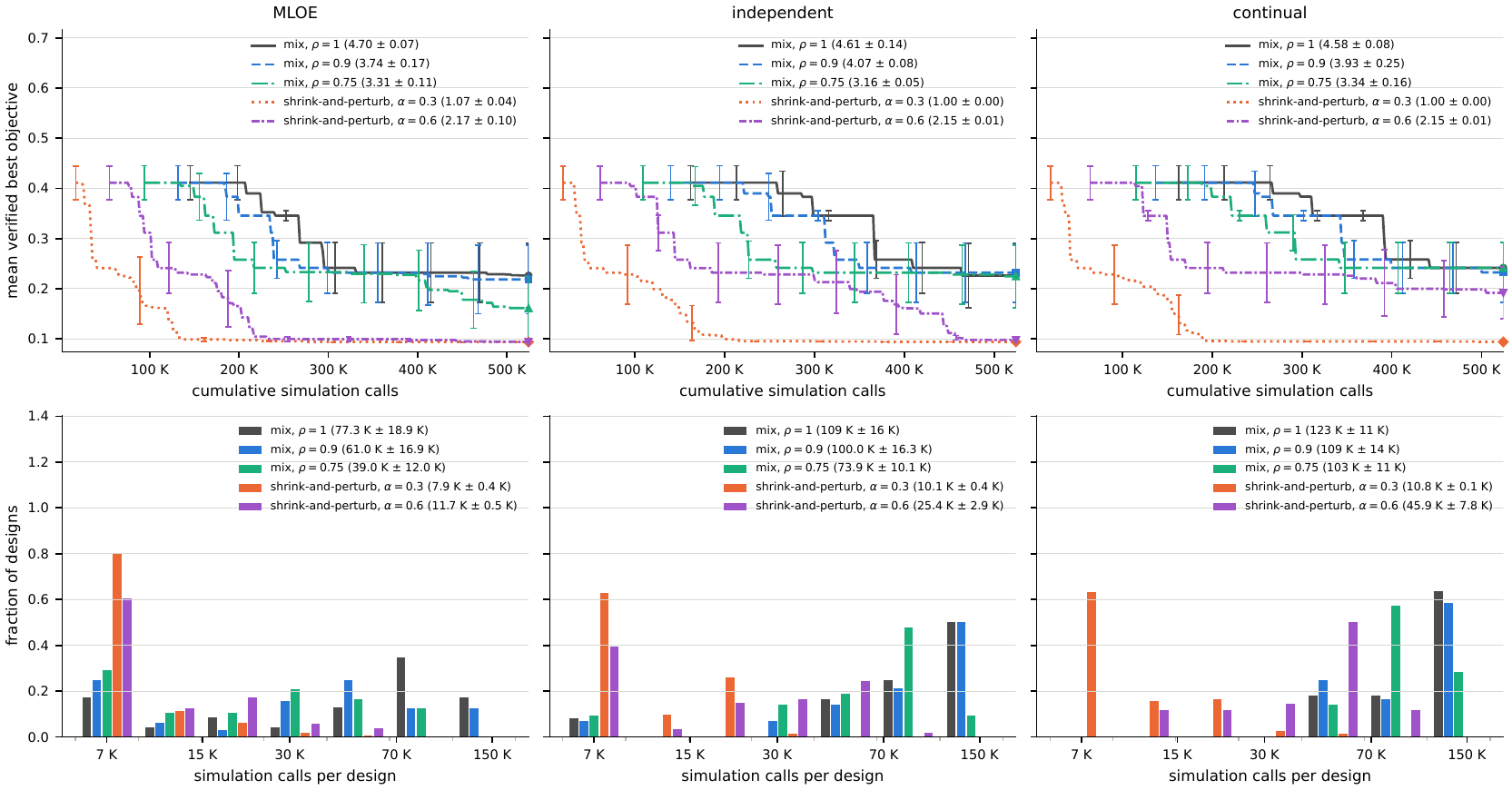}
    \caption{Warm-start training regimes on the probes for linear model, $m = 2$, one column per strategy. Top: verified best objective against cumulative simulation calls; average integral ranks are indicated in parentheses. Bottom: simulation calls per design; mean sample consumption per design is indicated in parentheses. Each warm-start training regime was evaluated on 3 independent runs.}
    \label{fig:regime-linear-2}
\end{figure}

\begin{figure}[p]
    \centering
    \includegraphics[width=\textwidth]{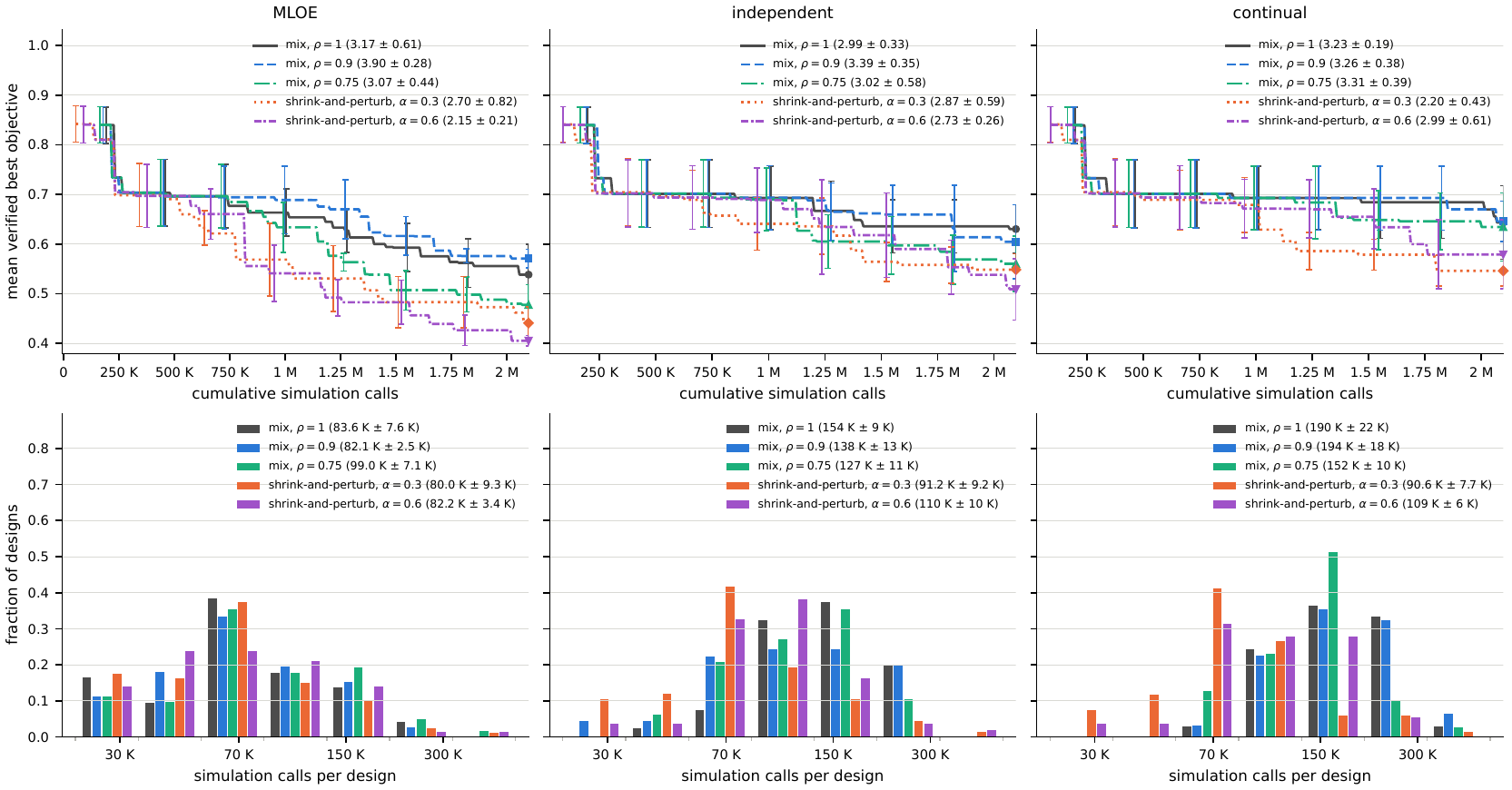}
    \caption{Warm-start training regimes on the enzyme inhibitor assay, one column per strategy. Top: verified best objective against cumulative simulation calls; average integral ranks are indicated in parentheses. Bottom: simulation calls per design; mean sample consumption per design is indicated in parentheses. Each warm-start training regime was evaluated on 3 independent runs.}
    \label{fig:regime-extremes}
\end{figure}

\begin{figure}[p]
    \centering
    \includegraphics[width=\textwidth]{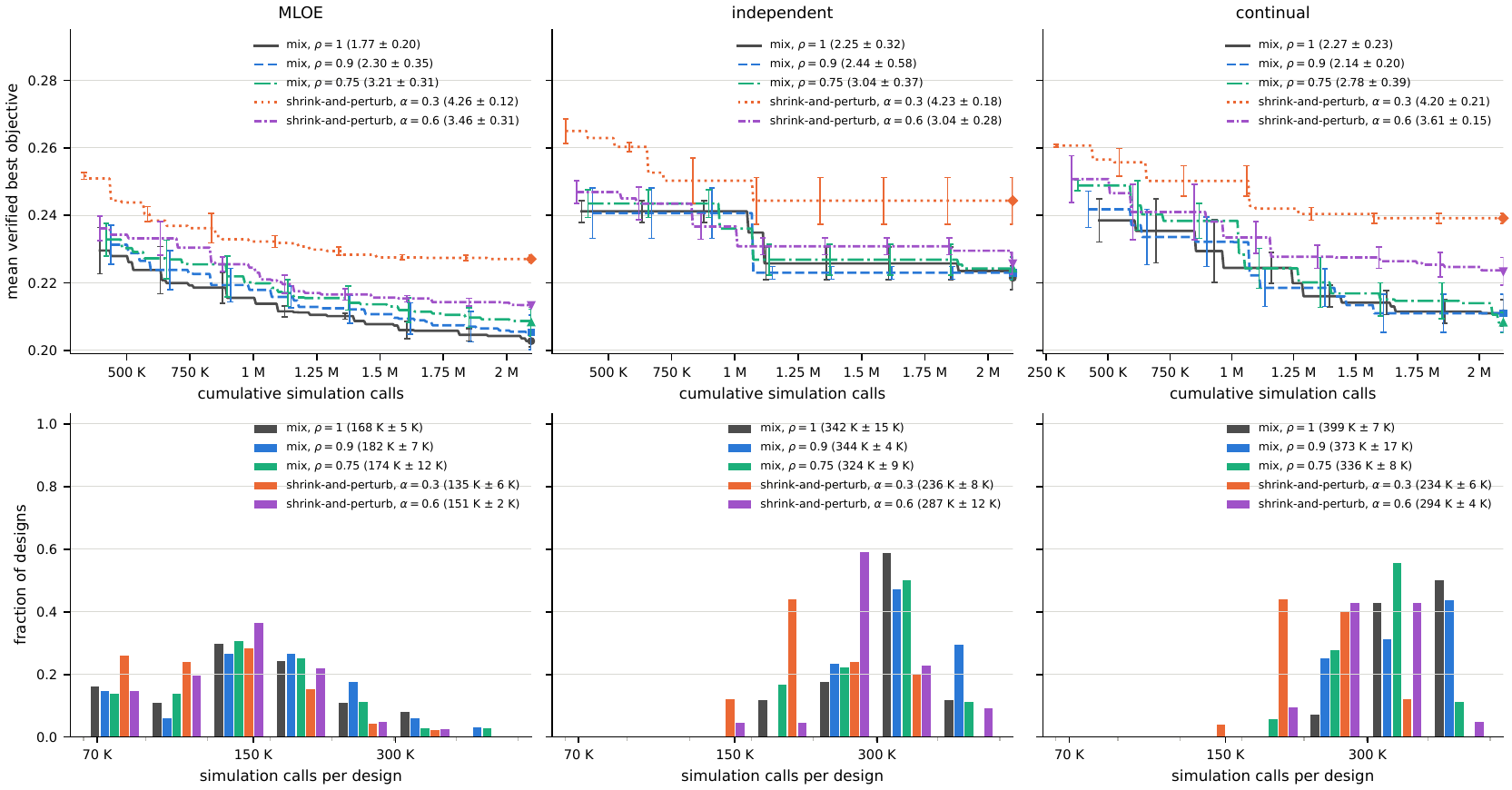}
    \caption{Warm-start training regimes on the SST angle task, one column per strategy. Top: verified best objective against cumulative simulation calls; average integral ranks are indicated in parentheses. Bottom: simulation calls per design; mean sample consumption per design is indicated in parentheses. Each warm-start training regime was evaluated on 3 independent runs.}
    \label{fig:regime-angle}
\end{figure}

\begin{figure}[p]
    \centering
    \includegraphics[width=\textwidth]{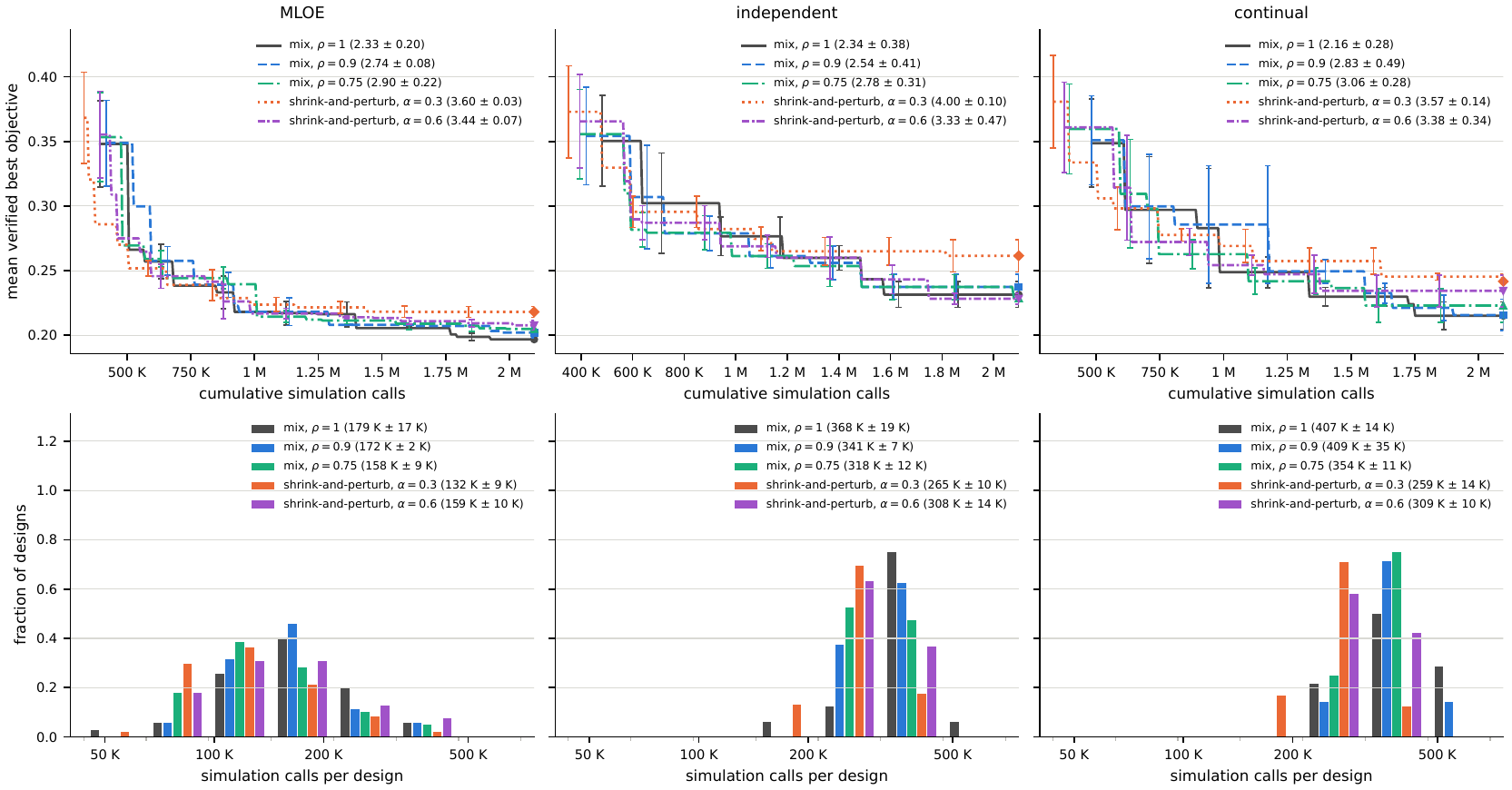}
    \caption{Warm-start training regimes on the SST layout task, one column per strategy. Top: verified best objective against cumulative simulation calls; average integral ranks are indicated in parentheses. Bottom: simulation calls per design; mean sample consumption per design is indicated in parentheses. Each warm-start training regime was evaluated on 3 independent runs.}
    \label{fig:regime-intersect}
\end{figure}

\begin{figure}[p]
    \centering
    \includegraphics[width=\textwidth]{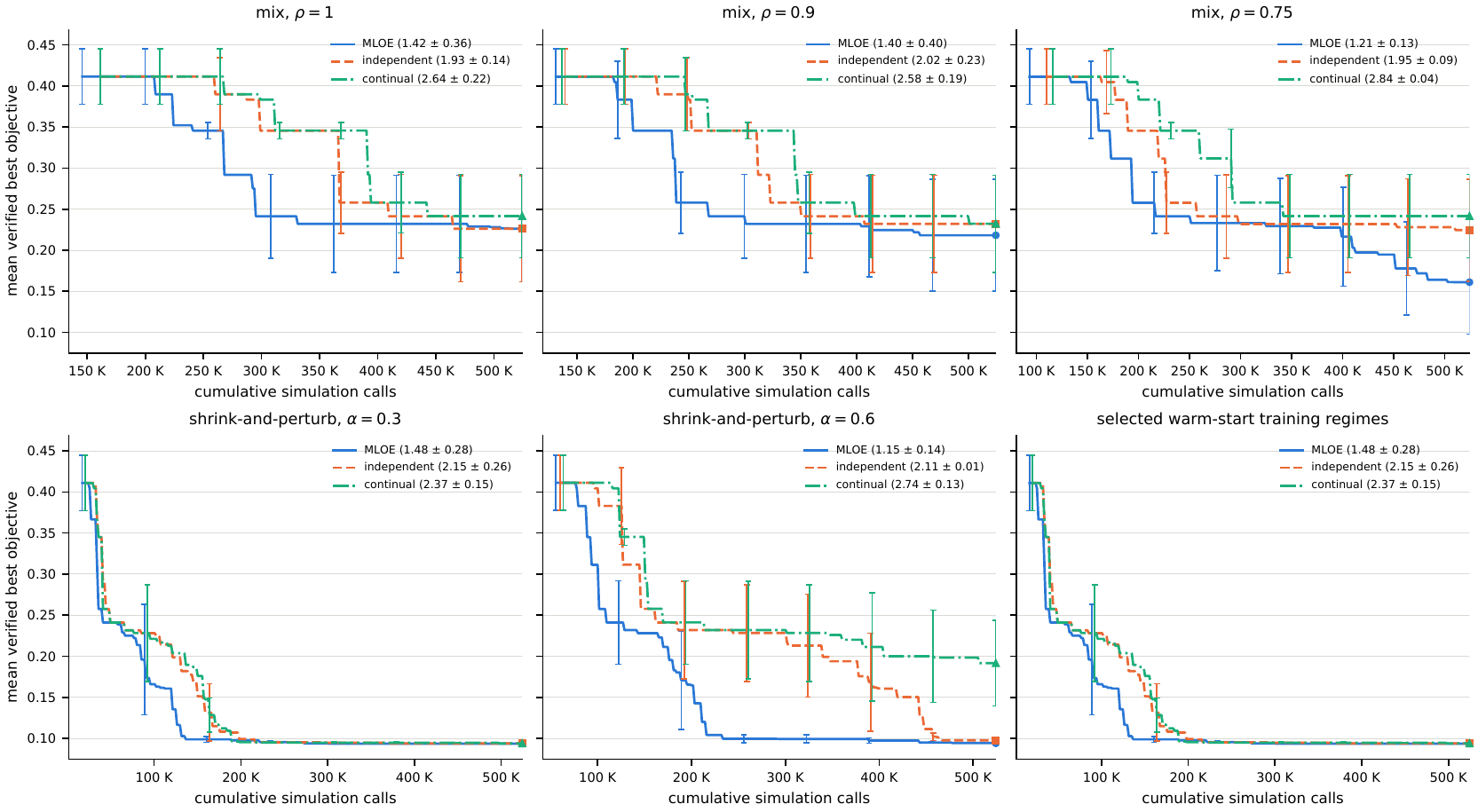}
    \caption{Training strategies under each warm-start training regime on the probes for linear model, $m = 2$, one panel per warm-start training regime; the last panel shows each strategy under the warm-start training regime selected for it. Verified best objective against cumulative simulation calls; average integral ranks are indicated in parentheses. Each strategy was evaluated on 3 independent runs.}
    \label{fig:strategy-by-regime-linear-2}
\end{figure}

\begin{figure}[p]
    \centering
    \includegraphics[width=\textwidth]{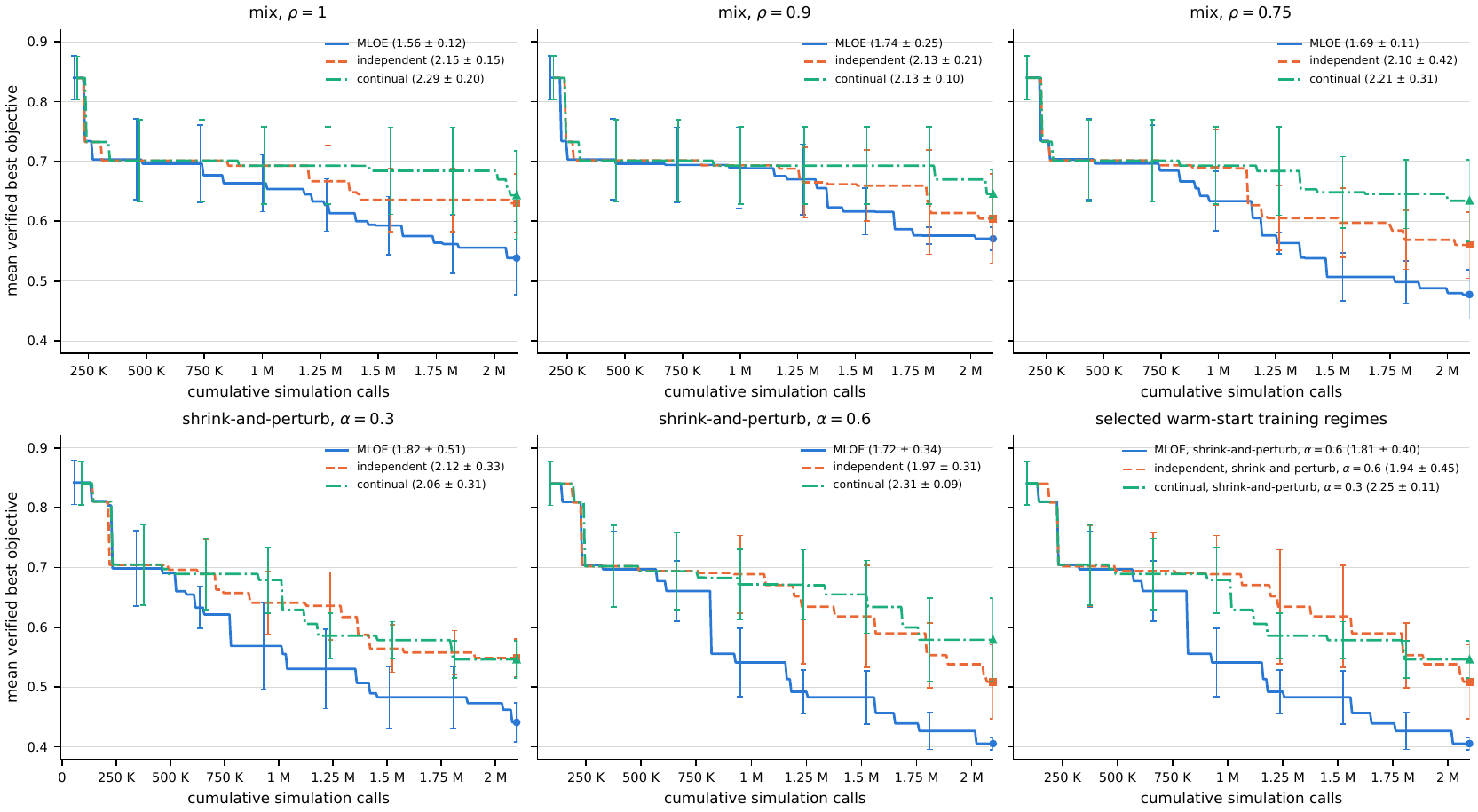}
    \caption{Training strategies under each warm-start training regime on the enzyme inhibitor assay, one panel per warm-start training regime; the last panel shows each strategy under the warm-start training regime selected for it. Verified best objective against cumulative simulation calls; average integral ranks are indicated in parentheses. Each strategy was evaluated on 3 independent runs.}
    \label{fig:strategy-by-regime-extremes}
\end{figure}

\begin{figure}[p]
    \centering
    \includegraphics[width=\textwidth]{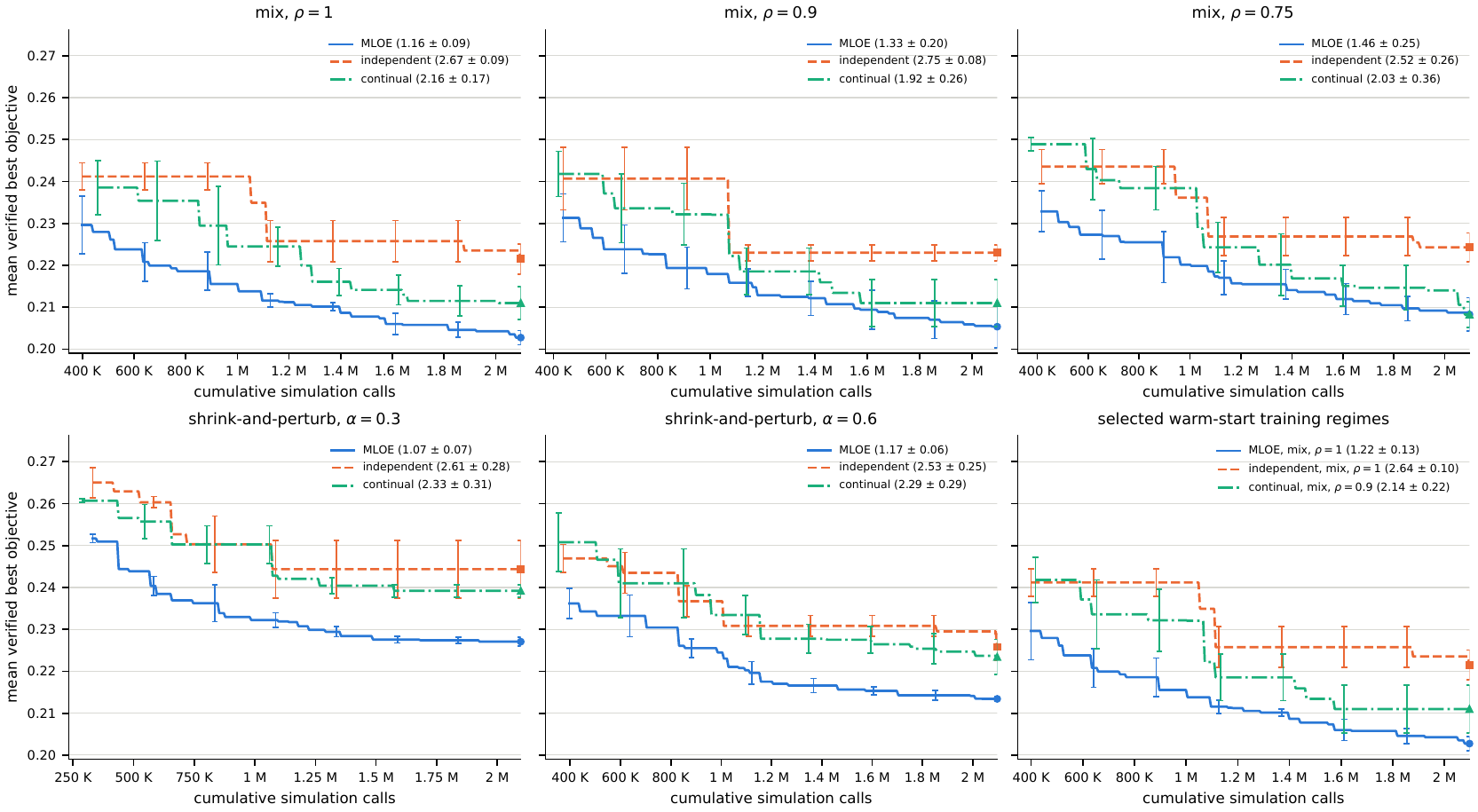}
    \caption{Training strategies under each warm-start training regime on the SST angle task, one panel per warm-start training regime; the last panel shows each strategy under the warm-start training regime selected for it. Verified best objective against cumulative simulation calls; average integral ranks are indicated in parentheses. Each strategy was evaluated on 3 independent runs.}
    \label{fig:strategy-by-regime-angle}
\end{figure}

\begin{figure}[p]
    \centering
    \includegraphics[width=\textwidth]{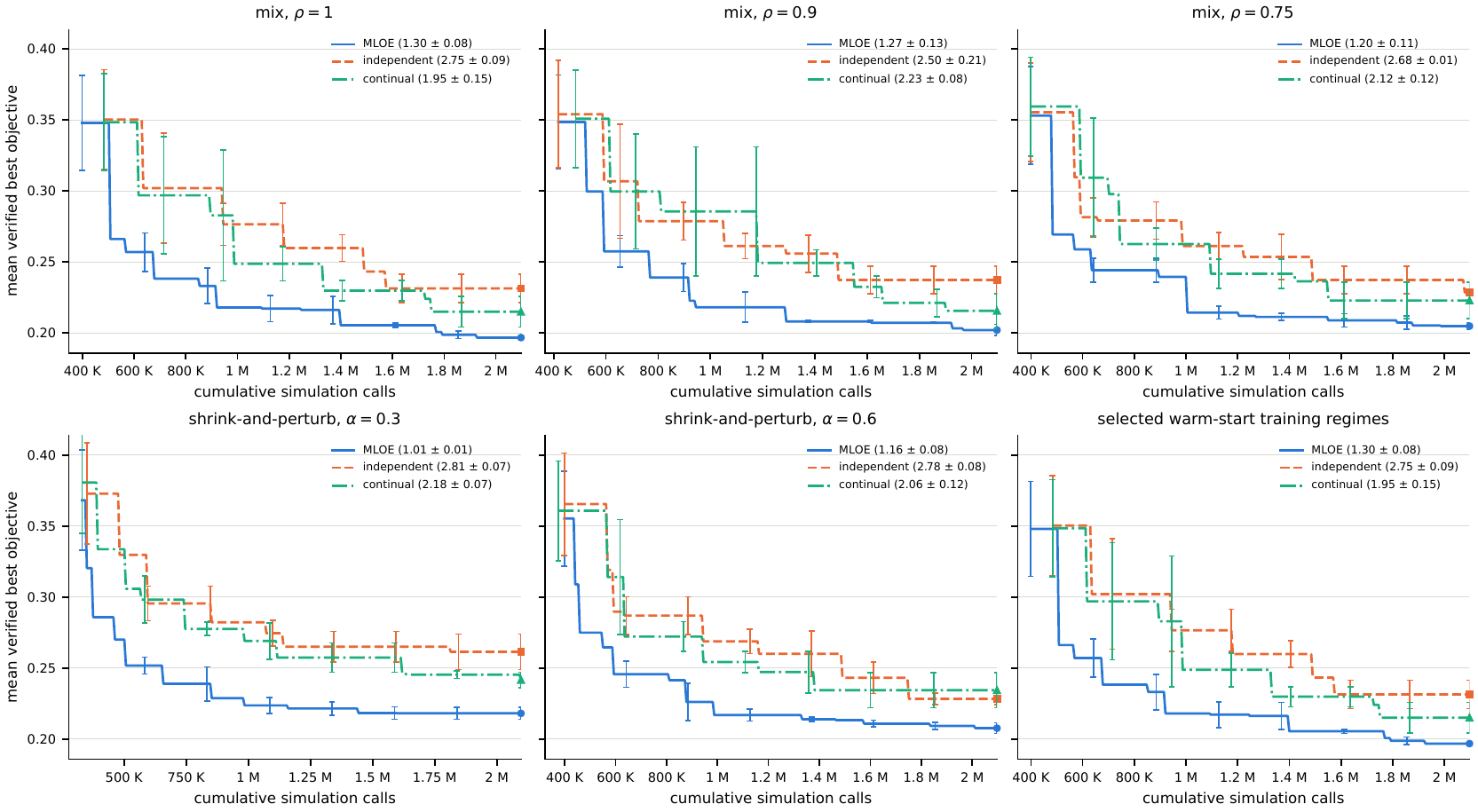}
    \caption{Training strategies under each warm-start training regime on the SST layout task, one panel per warm-start training regime; the last panel shows each strategy under the warm-start training regime selected for it. Verified best objective against cumulative simulation calls; average integral ranks are indicated in parentheses. Each strategy was evaluated on 3 independent runs.}
    \label{fig:strategy-by-regime-intersect}
\end{figure}

Results of hyperparameter selection are illustrated in figures~\ref{fig:regime-linear-2} to~\ref{fig:regime-intersect} and summarized in table~\ref{tab:selection}. Additionally, figures~\ref{fig:strategy-by-regime-linear-2} to~\ref{fig:strategy-by-regime-intersect} compare the strategies under each warm-start training regime, and table~\ref{tab:rank-by-regime} lists the average integral ranks. Results show that the best warm-start training regime changes depending on the task. The findings of Ash and Adams~\cite{ash2020warm} suggest that the best regime also depends on the network architecture. For practical applications, we recommend running a small-scale comparison between warm-start training regimes on a few detector designs before committing to the proper optimization run. We also note that MLOE converges faster than the baseline strategies under any fixed warm-start training regime.

\begin{table}[!ht]
    \caption{Results of hyperparameter optimization: average integral rank of each warm-start training regime for each training strategy, computed on the verified objective values over the three selection runs, mean $\pm$ standard error; on the probes for linear model, one selection is pooled over the four dimensionalities, twelve runs per strategy. Lower is better; 1 is the best attainable and 5 the worst. The lowest values are marked in bold, the values within two combined standard errors~---~in italics.}
    \centering
    \begin{tabular}{@{}llccc@{}}
        \toprule
        Task & Warm-start training regime & MLOE & independent & continual \\
        \midrule
        probes for linear model & mix, $\rho = 1$ (none) & $4.53 \pm 0.08$ & $4.65 \pm 0.06$ & $4.60 \pm 0.07$ \\
         & mix, $\rho = 0.9$ & $4.08 \pm 0.09$ & $4.09 \pm 0.05$ & $4.10 \pm 0.09$ \\
         & mix, $\rho = 0.75$ & $3.20 \pm 0.06$ & $3.11 \pm 0.03$ & $3.17 \pm 0.05$ \\
         & shrink-and-perturb, $\alpha = 0.3$ & $\bm{1.23 \pm 0.12}$ & $\bm{1.05 \pm 0.05}$ & $\bm{1.04 \pm 0.04}$ \\
         & shrink-and-perturb, $\alpha = 0.6$ & $1.96 \pm 0.10$ & $2.10 \pm 0.03$ & $2.09 \pm 0.05$ \\
        \midrule
        enzyme inhibitor assay & mix, $\rho = 1$ (none) & \textit{3.17 $\pm$ 0.61} & \textit{2.99 $\pm$ 0.33} & $3.23 \pm 0.19$ \\
         & mix, $\rho = 0.9$ & $3.90 \pm 0.28$ & \textit{3.39 $\pm$ 0.35} & \textit{3.26 $\pm$ 0.38} \\
         & mix, $\rho = 0.75$ & \textit{3.07 $\pm$ 0.44} & \textit{3.02 $\pm$ 0.58} & \textit{3.31 $\pm$ 0.39} \\
         & shrink-and-perturb, $\alpha = 0.3$ & \textit{2.70 $\pm$ 0.82} & \textit{2.87 $\pm$ 0.59} & $\bm{2.20 \pm 0.43}$ \\
         & shrink-and-perturb, $\alpha = 0.6$ & $\bm{2.15 \pm 0.21}$ & $\bm{2.73 \pm 0.26}$ & \textit{2.99 $\pm$ 0.61} \\
        \midrule
        SST angle & mix, $\rho = 1$ (none) & $\bm{1.77 \pm 0.20}$ & $\bm{2.25 \pm 0.32}$ & \textit{2.27 $\pm$ 0.23} \\
         & mix, $\rho = 0.9$ & \textit{2.30 $\pm$ 0.35} & \textit{2.44 $\pm$ 0.58} & $\bm{2.14 \pm 0.20}$ \\
         & mix, $\rho = 0.75$ & $3.21 \pm 0.31$ & \textit{3.04 $\pm$ 0.37} & \textit{2.78 $\pm$ 0.39} \\
         & shrink-and-perturb, $\alpha = 0.3$ & $4.26 \pm 0.12$ & $4.23 \pm 0.18$ & $4.20 \pm 0.21$ \\
         & shrink-and-perturb, $\alpha = 0.6$ & $3.46 \pm 0.31$ & \textit{3.04 $\pm$ 0.28} & $3.61 \pm 0.15$ \\
        \midrule
        SST layout & mix, $\rho = 1$ (none) & $\bm{2.33 \pm 0.20}$ & $\bm{2.34 \pm 0.38}$ & $\bm{2.16 \pm 0.28}$ \\
         & mix, $\rho = 0.9$ & \textit{2.74 $\pm$ 0.08} & \textit{2.54 $\pm$ 0.41} & \textit{2.83 $\pm$ 0.49} \\
         & mix, $\rho = 0.75$ & \textit{2.90 $\pm$ 0.22} & \textit{2.78 $\pm$ 0.31} & $3.06 \pm 0.28$ \\
         & shrink-and-perturb, $\alpha = 0.3$ & $3.60 \pm 0.03$ & $4.00 \pm 0.10$ & $3.57 \pm 0.14$ \\
         & shrink-and-perturb, $\alpha = 0.6$ & $3.44 \pm 0.07$ & \textit{3.33 $\pm$ 0.47} & $3.38 \pm 0.34$ \\
\bottomrule
    \end{tabular}
    \label{tab:selection}
\end{table}

\begin{table}[!ht]
    \caption{Average integral rank of each training strategy under each warm-start training regime ($\rho$: mixing, $\alpha$: shrink-and-perturb), on the verified objective values over three seeds, mean $\pm$ standard error; on the probes for linear model, over the three seeds of each of the four dimensionalities. Lower is better; 1 is the best attainable and 3 the worst. The lowest values are marked in bold, the values within two combined standard errors~---~in italics.}
    \centering
    \widetablesetup
    \begin{tabular}{@{}llccccc@{}}
        \toprule
        Task & Strategy & $\rho = 1$ & $\rho = 0.9$ & $\rho = 0.75$ & $\alpha = 0.3$ & $\alpha = 0.6$ \\
        \midrule
        probes for linear model & MLOE & $\bm{1.17 \pm 0.10}$ & $\bm{1.27 \pm 0.12}$ & $\bm{1.23 \pm 0.09}$ & $\bm{1.63 \pm 0.17}$ & $\bm{1.08 \pm 0.05}$ \\
         & independent & $2.26 \pm 0.12$ & $2.04 \pm 0.10$ & $1.91 \pm 0.09$ & $2.11 \pm 0.16$ & $2.37 \pm 0.10$ \\
         & continual & $2.57 \pm 0.10$ & $2.69 \pm 0.08$ & $2.86 \pm 0.02$ & $2.26 \pm 0.11$ & $2.56 \pm 0.09$ \\
        \midrule
        enzyme inhibitor assay & MLOE & $\bm{1.56 \pm 0.12}$ & $\bm{1.74 \pm 0.25}$ & $\bm{1.69 \pm 0.11}$ & $\bm{1.82 \pm 0.51}$ & $\bm{1.72 \pm 0.34}$ \\
         & independent & $2.15 \pm 0.15$ & \textit{2.13 $\pm$ 0.21} & \textit{2.10 $\pm$ 0.42} & \textit{2.12 $\pm$ 0.33} & \textit{1.97 $\pm$ 0.31} \\
         & continual & $2.29 \pm 0.20$ & \textit{2.13 $\pm$ 0.10} & \textit{2.21 $\pm$ 0.31} & \textit{2.06 $\pm$ 0.31} & \textit{2.31 $\pm$ 0.09} \\
        \midrule
        SST angle & MLOE & $\bm{1.16 \pm 0.09}$ & $\bm{1.33 \pm 0.20}$ & $\bm{1.46 \pm 0.25}$ & $\bm{1.07 \pm 0.07}$ & $\bm{1.17 \pm 0.06}$ \\
         & independent & $2.67 \pm 0.09$ & $2.75 \pm 0.08$ & $2.52 \pm 0.26$ & $2.61 \pm 0.28$ & $2.53 \pm 0.25$ \\
         & continual & $2.16 \pm 0.17$ & \textit{1.92 $\pm$ 0.26} & \textit{2.03 $\pm$ 0.36} & $2.33 \pm 0.31$ & $2.29 \pm 0.29$ \\
        \midrule
        SST layout & MLOE & $\bm{1.30 \pm 0.08}$ & $\bm{1.27 \pm 0.13}$ & $\bm{1.20 \pm 0.11}$ & $\bm{1.01 \pm 0.01}$ & $\bm{1.16 \pm 0.08}$ \\
         & independent & $2.75 \pm 0.09$ & $2.50 \pm 0.21$ & $2.68 \pm 0.01$ & $2.81 \pm 0.07$ & $2.78 \pm 0.08$ \\
         & continual & $1.95 \pm 0.15$ & $2.23 \pm 0.08$ & $2.12 \pm 0.12$ & $2.18 \pm 0.07$ & $2.06 \pm 0.12$ \\
\bottomrule
    \end{tabular}
    \label{tab:rank-by-regime}
\end{table}

\section{Network architectures}\label{app:architecture}

\begin{figure}
    \centering
    \includegraphics[width=\textwidth]{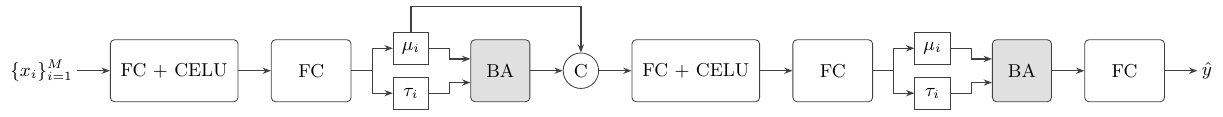}
    \caption{Architecture for the inference model used in all experiments. The fully connected (FC) layers act on every element;  each Bayesian aggregation (BA) combines the per-element means $\mu_i$ and precisions $\tau_i$ across elements; the circled C denotes concatenation of that representation with the mean of every element.}
    \label{fig:architecture}
\end{figure}

The inference tasks in all experiments are permutation-invariant with respect to the inputs. For all experiments, we employ the same permutation-invariant network based on Deep Sets~\cite{zaheer2017deep}. The architecture is shown in figure~\ref{fig:architecture}. An event is a set of $M$ elements (probes, experiments or hits), each carrying a feature vector. The network has two blocks, each block has two fully connected layers, followed by the aggregation. Block is applied to element-wise, aggregation combines per-element features, results of intermediate aggregations are concatenated with pre-aggregation features of each element.
In the selected experiments, elements might not carry equally important information, we choose Bayesian aggregation~\cite{volpp2021bayesian} (BA) to express this notion explicitly. BA combines elements' features $\mu_i$ weighted by precision $\tau_i > 0$:
\begin{align*}
    e = \frac{\sum_i \tau_i \odot \mu_i}{\sum_i \tau_i + 1},
\end{align*}
where both quantities are produced by the preceding layer.

The meta-inference model and the baseline models share this architecture and differ only in the per-element input as described in the section~\ref{sec:experiments}. The widths, the targets and the resulting parameter counts are listed in table~\ref{tab:architecture}. On the linear and enzyme tasks the regressor is an ensemble of four independent networks evaluated in a single batched pass, in the manner of TabM~\cite{gorishniy2025tabm}; the members are trained on independent mini-batches and their predictions are averaged at evaluation, which avoids the learner collusion that a joint loss on the averaged prediction induces~\cite{jeffares2023joint}.

The networks are trained with AdamW~\cite{loshchilov2018decoupled} ($\beta_1 = 0.9$, $\beta_2 = 0.999$, weight decay $10^{-3}$) at a base learning rate of $5 \times 10^{-4}$ on the SST tasks and $2.5 \times 10^{-4}$ on the linear and enzyme tasks, with mini-batches of 256 events on the SST and enzyme tasks and 128 on the linear task. To accelerate training after each data addition, the learning rate follows a cosine schedule~\cite{loshchilov2017sgdr} that is restarted at every acquisition, including the initial one: it decays from four times the base learning rate to the base value over 16 epochs and remains at the base value afterwards.

\begin{table}[!ht]
    \caption{Parameters of the inference model: per-element input width for MLOE and for the baselines, hidden and output widths of the two blocks, target width, and parameter count of one network. Ensembles have four identical networks.}
    \centering
    \begin{tabular}{@{}lccccc@{}}
        \toprule
        Task & Input (MLOE / baseline) & Block 1 & Block 2 & Target & Parameters \\
        \midrule
        probes for linear model, $m$ & $1 + m$ / $2 + m$ & $24 \to 32$ & $32 \to 24$ & $m + 1$ & 5.4--5.6 $\times 10^3$ \\
        enzyme inhibitor assay & $36$ / $36$ & $24 \to 16$ & $16 \to 24$ & $2$ & $3.1 \times 10^3$ \\
        SST angle, SST layout & $4$ / $5$ & $32 \to 48$ & $48 \to 32$ & $9$ & $1.14 \times 10^4$ \\
        \bottomrule
    \end{tabular}
    \label{tab:architecture}
\end{table}

\section{Weight of the context samples}\label{app:lambda}
In this study, we investigate effects of weight $\lambda$ of the context samples in equation~\eqref{eq:structured-estimate}.
For all tasks, we perform a full reproduction of the main results under $\lambda = 1/4$ (the main experiments use $\lambda = 1$); for both SST tasks we additionally probe $\lambda = 1/8$ and $\lambda = 1/2$ on the warm-start training regime selection seeds.

\begin{figure}[p]
    \centering
    \includegraphics[width=\textwidth]{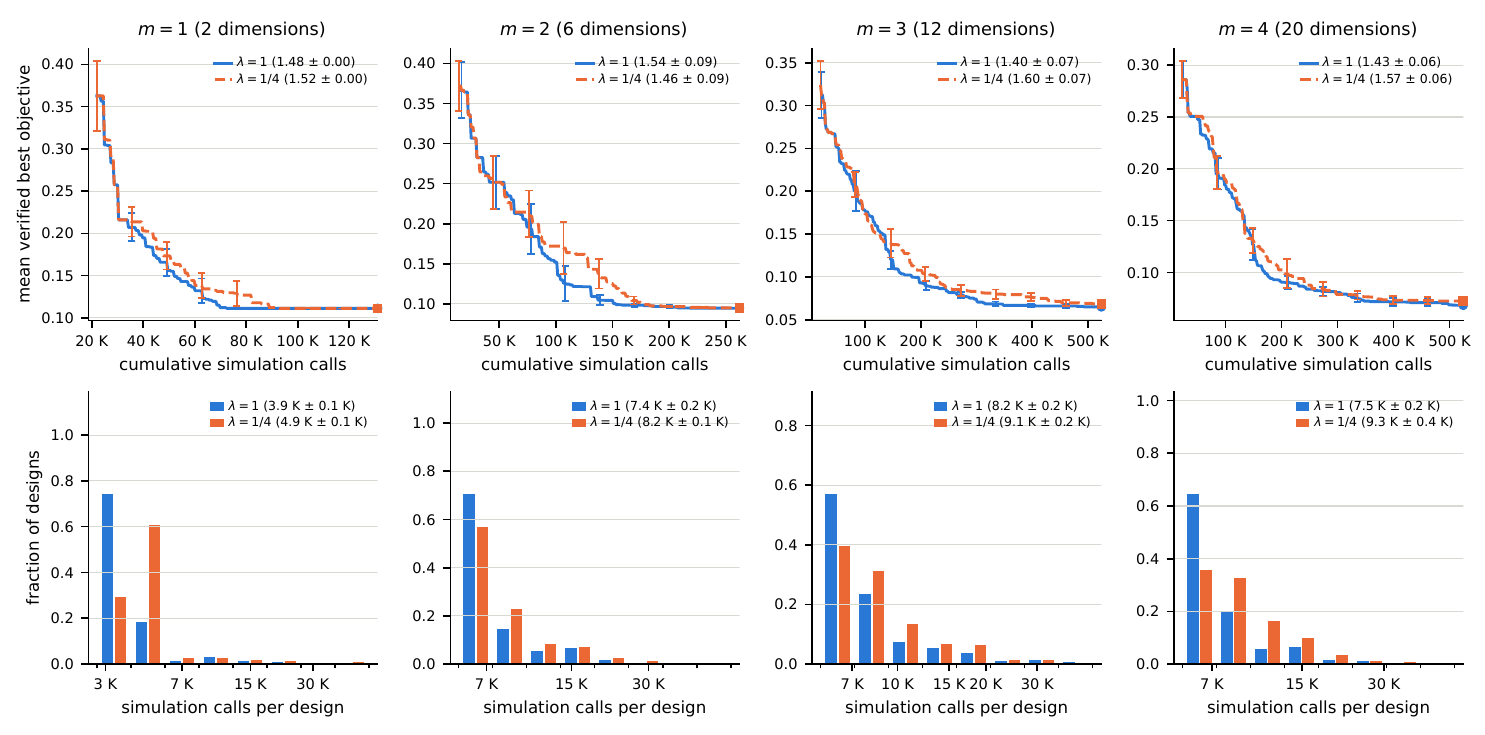}
    \caption{MLOE with the weight $\lambda$ of the context samples varied, on the probes for linear model, one column per dimensionality $m$. Top: verified best objective against cumulative simulation calls; average integral ranks are indicated in parentheses. Bottom: simulation calls per design; mean sample consumption per design is indicated in parentheses. Ten seeds.
    For illustration purposes, convergence curves for $m = 1$ and $m = 2$ are cut to $2^{17}$ and $2^{18}$ simulation calls.}
    \label{fig:ablation-lambda-linear}
\end{figure}

\begin{figure}[p]
    \centering
    \includegraphics[width=\textwidth]{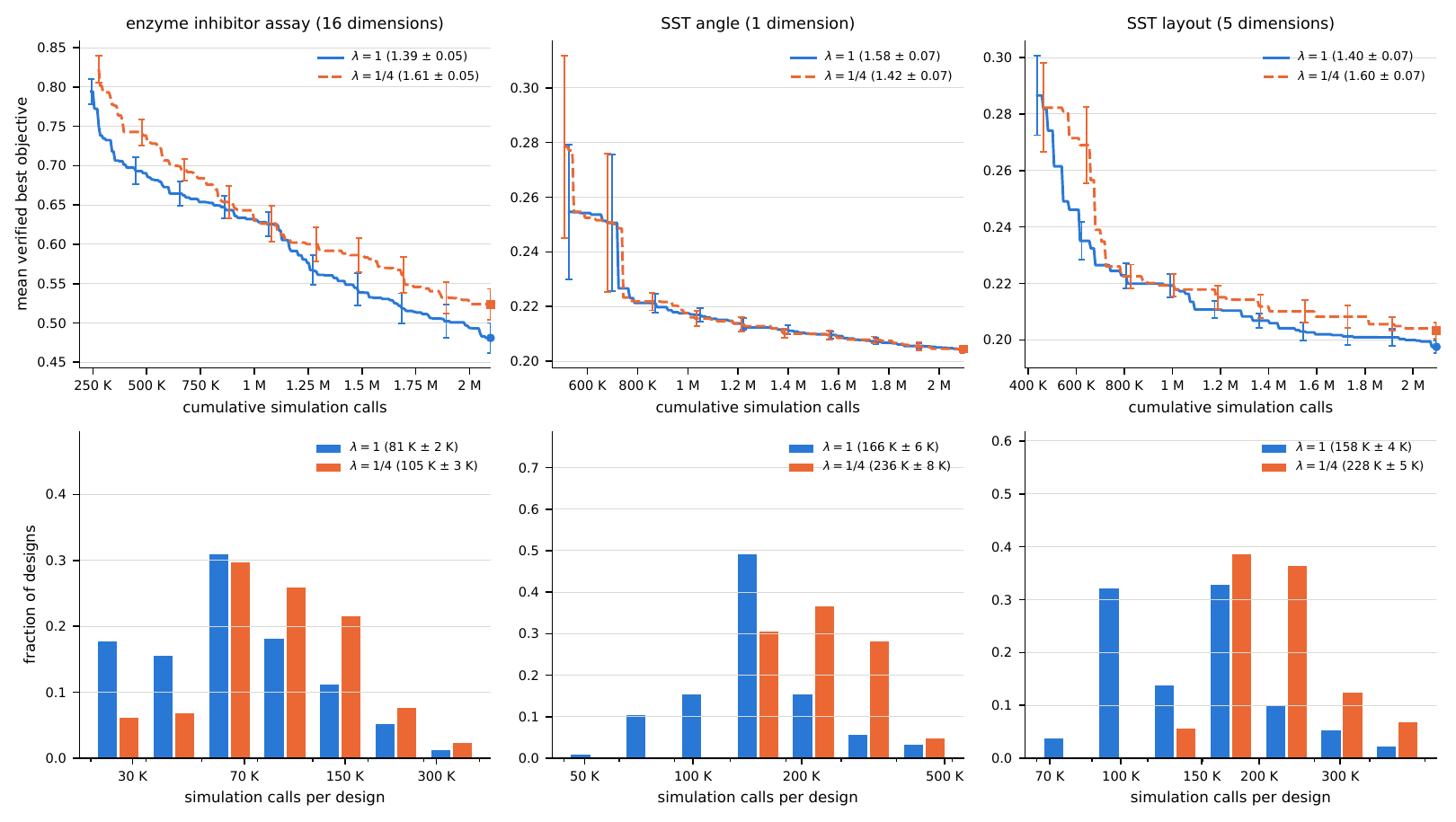}
    \caption{MLOE with the weight $\lambda$ of the context samples varied, on the enzyme inhibitor assay (left), SST angle (middle) and SST layout (right) tasks. Top: verified best objective against cumulative simulation calls; average integral ranks are indicated in parentheses. Bottom: simulation calls per design; mean sample consumption per design is indicated in parentheses. Ten seeds of the test runs per SST task, twenty on the enzyme inhibitor assay.}
    \label{fig:ablation-lambda}
\end{figure}

Figures~\ref{fig:ablation-lambda-linear}~and~\ref{fig:ablation-lambda} show the convergence curves and the sample consumption, and tables~\ref{tab:rank-lambda-linear}~and~\ref{tab:rank-lambda} list the average integral ranks and the sample consumption: MLOE with different values of $\lambda$ against the baseline strategies, and the values of $\lambda$ among themselves.
The weight $\lambda$ does not alter the ranking against the baseline methods on any task except the enzyme inhibitor assay, where MLOE is worse than the baseline methods; however, all strategies are within two standard errors of each other. The comparison across the values of $\lambda$ reveals that the sample consumption per design decreases monotonically with $\lambda$ on all tasks. On most of the tasks, MLOE with $\lambda = 1$ is either better than or on par with $\lambda = 1 / 4$.

\begin{figure}[p]
    \centering
    \includegraphics[width=\textwidth]{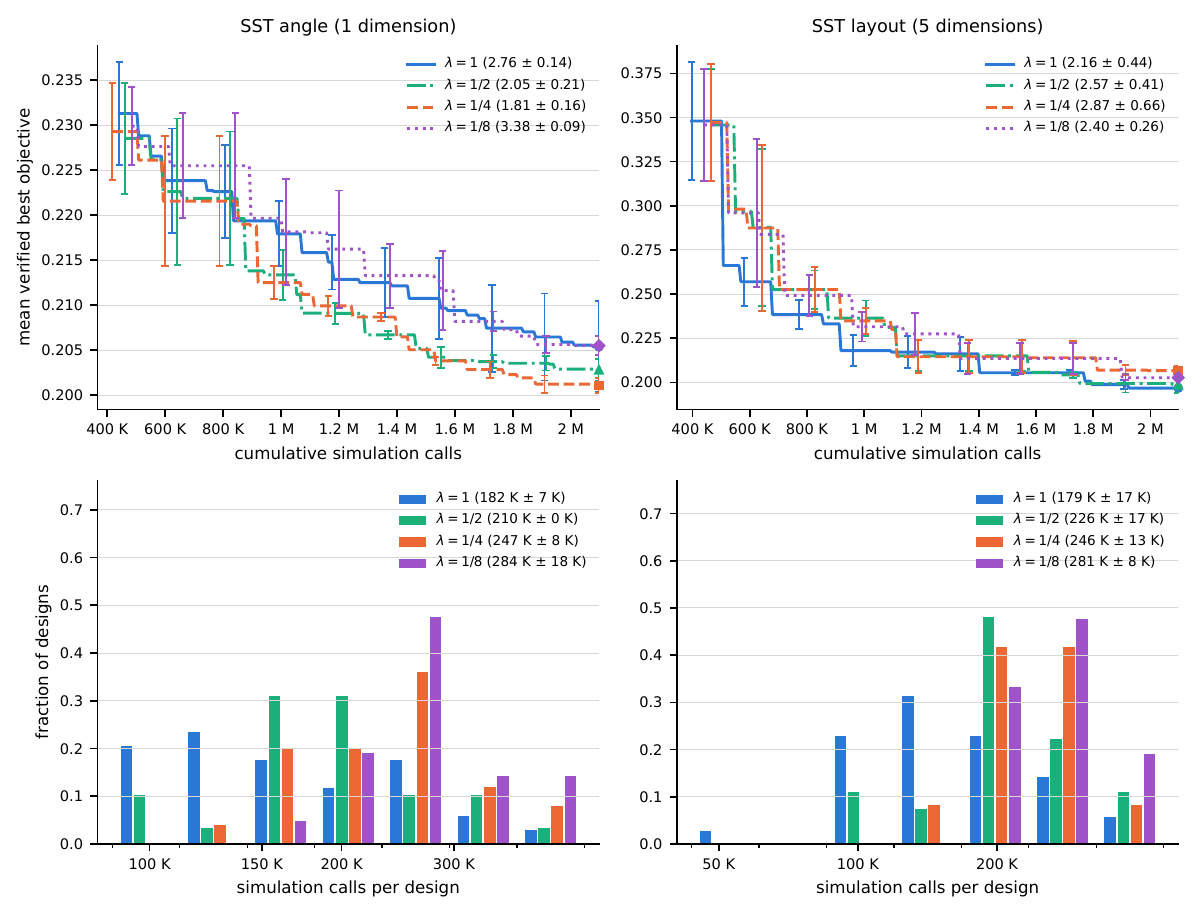}
    \caption{MLOE with the weight $\lambda$ of the context samples varied over $1$, $1/2$, $1/4$ and $1/8$, on the SST angle (left) and SST layout (right) tasks, at \emph{a common warm-start training regime}. Top: verified best objective against cumulative simulation calls; average integral ranks are indicated in parentheses. Bottom: simulation calls per design; mean sample consumption per design is indicated in parentheses. Three seeds of the selection runs.}
    \label{fig:ablation-lambda-scan}
\end{figure}

Evaluations on a finer $\lambda$ grid, shown in figure~\ref{fig:ablation-lambda-scan}~and~table~\ref{tab:rank-lambda-scan}, further support the findings: with decreasing $\lambda$, the sample consumption of MLOE increases, approaching that of the baseline strategies. The unusual preference towards $\lambda = 1/4$ and $\lambda = 1/2$ on the SST angle task is, most likely, a result of low statistics: the finer grid was evaluated on three seeds, while a larger sample (figure~\ref{fig:ablation-lambda}, middle column) shows that $\lambda = 1$ and $\lambda = 1/4$ are statistically indistinguishable.

\begin{table}[!ht]
    \caption{Average integral rank and mean simulation calls per design on the ten seeds of the test runs of the probes for linear model, mean $\pm$ standard error. First two blocks: MLOE with the given weight $\lambda$ against the baseline strategies; last block: the two values of $\lambda$ ranked between themselves. Lower is better; 1 is the best attainable, 3 the worst in the first two blocks and 2 in the last. The lowest values are marked in bold, the values within two combined standard errors~---~in italics.}
    \centering
    \footnotesize
    \setlength{\tabcolsep}{1.5pt}
    \begin{tabular}{@{}llcccccccc@{}}
        \toprule
        & & \multicolumn{2}{c}{$m = 1$} & \multicolumn{2}{c}{$m = 2$} & \multicolumn{2}{c}{$m = 3$} & \multicolumn{2}{c}{$m = 4$} \\
        \cmidrule(lr){3-4} \cmidrule(lr){5-6} \cmidrule(lr){7-8} \cmidrule(lr){9-10}
        & Strategy & rank & calls, $10^3$ & rank & calls, $10^3$ & rank & calls, $10^3$ & rank & calls, $10^3$ \\
        \midrule
        $\lambda = 1$ & MLOE & $\bm{1.92 \pm 0.01}$ & $\bm{3.9 \pm 0.1}$ & $\bm{1.53 \pm 0.10}$ & $\bm{7.4 \pm 0.2}$ & $\bm{1.32 \pm 0.06}$ & $\bm{8.2 \pm 0.2}$ & $\bm{1.54 \pm 0.13}$ & $\bm{7.5 \pm 0.2}$ \\
         & independent & $2.04 \pm 0.01$ & $5.5 \pm 0.1$ & $2.26 \pm 0.13$ & $10.5 \pm 0.3$ & $2.46 \pm 0.09$ & $13.0 \pm 0.6$ & $2.19 \pm 0.10$ & $13.4 \pm 0.5$ \\
         & continual & $2.04 \pm 0.01$ & $5.5 \pm 0.2$ & $2.21 \pm 0.12$ & $10.4 \pm 0.1$ & $2.22 \pm 0.11$ & $11.6 \pm 0.4$ & $2.28 \pm 0.15$ & $11.3 \pm 0.5$ \\
        \midrule
        $\lambda = 1/4$ & MLOE & $\bm{1.95 \pm 0.02}$ & $\bm{4.9 \pm 0.1}$ & $\bm{1.69 \pm 0.16}$ & $\bm{8.2 \pm 0.1}$ & $\bm{1.45 \pm 0.13}$ & $\bm{9.1 \pm 0.2}$ & $\bm{1.66 \pm 0.10}$ & $\bm{9.3 \pm 0.4}$ \\
         & independent & $2.03 \pm 0.02$ & $5.5 \pm 0.1$ & $2.21 \pm 0.13$ & $10.5 \pm 0.3$ & $2.42 \pm 0.11$ & $13.0 \pm 0.6$ & $2.13 \pm 0.12$ & $13.4 \pm 0.5$ \\
         & continual & $2.03 \pm 0.01$ & $5.5 \pm 0.2$ & $2.10 \pm 0.11$ & $10.4 \pm 0.1$ & $2.13 \pm 0.11$ & $11.6 \pm 0.4$ & $2.22 \pm 0.13$ & $11.3 \pm 0.5$ \\
        \midrule
        across $\lambda$ & $\lambda = 1$ & $\bm{1.48 \pm 0.00}$ & $\bm{3.9 \pm 0.1}$ & \textit{1.54 $\pm$ 0.09} & $\bm{7.4 \pm 0.2}$ & $\bm{1.40 \pm 0.07}$ & $\bm{8.2 \pm 0.2}$ & $\bm{1.43 \pm 0.06}$ & $\bm{7.5 \pm 0.2}$ \\
         & $\lambda = 1/4$ & $1.52 \pm 0.00$ & $4.9 \pm 0.1$ & $\bm{1.46 \pm 0.09}$ & $8.2 \pm 0.1$ & $1.60 \pm 0.07$ & $9.1 \pm 0.2$ & \textit{1.57 $\pm$ 0.06} & $9.3 \pm 0.4$ \\
\bottomrule
    \end{tabular}
    \label{tab:rank-lambda-linear}
\end{table}

\begin{table}[!ht]
    \caption{Average integral rank and mean simulation calls per design on the ten seeds of the test runs of the SST tasks and the twenty seeds of the test runs of the enzyme inhibitor assay, mean $\pm$ standard error. First two blocks: MLOE with the given weight $\lambda$ against the baseline strategies; last block: the two values of $\lambda$ ranked between themselves. Lower is better; 1 is the best attainable, 3 the worst in the first two blocks and 2 in the last. The lowest values are marked in bold, the values within two combined standard errors~---~in italics.}
    \centering
    \footnotesize
    \setlength{\tabcolsep}{3pt}
    \begin{tabular}{@{}llcccccc@{}}
        \toprule
        & & \multicolumn{2}{c}{enzyme inhibitor assay} & \multicolumn{2}{c}{SST angle} & \multicolumn{2}{c}{SST layout} \\
        \cmidrule(lr){3-4} \cmidrule(lr){5-6} \cmidrule(lr){7-8}
        & Strategy & rank & calls, $10^3$ & rank & calls, $10^3$ & rank & calls, $10^3$ \\
        \midrule
        $\lambda = 1$ & MLOE & $\bm{1.83 \pm 0.08}$ & $\bm{81.2 \pm 1.8}$ & $\bm{1.34 \pm 0.09}$ & $\bm{165.6 \pm 5.7}$ & $\bm{1.14 \pm 0.06}$ & $\bm{157.9 \pm 4.4}$ \\
         & independent & $2.16 \pm 0.08$ & $105.9 \pm 2.6$ & $2.58 \pm 0.10$ & $317.6 \pm 7.3$ & $2.72 \pm 0.06$ & $324.5 \pm 6.4$ \\
         & continual & \textit{2.01 $\pm$ 0.10} & $90.0 \pm 2.3$ & $2.07 \pm 0.13$ & $342.6 \pm 9.8$ & $2.14 \pm 0.08$ & $360.8 \pm 8.9$ \\
        \midrule
        $\lambda = 1/4$ & MLOE & \textit{2.10 $\pm$ 0.08} & $104.8 \pm 3.1$ & $\bm{1.30 \pm 0.11}$ & $\bm{236.0 \pm 7.7}$ & $\bm{1.25 \pm 0.09}$ & $\bm{227.6 \pm 4.6}$ \\
         & independent & \textit{2.02 $\pm$ 0.10} & $105.9 \pm 2.6$ & $2.61 \pm 0.09$ & $317.6 \pm 7.3$ & $2.62 \pm 0.06$ & $324.5 \pm 6.4$ \\
         & continual & $\bm{1.88 \pm 0.10}$ & $\bm{90.0 \pm 2.3}$ & $2.08 \pm 0.15$ & $342.6 \pm 9.8$ & $2.13 \pm 0.07$ & $360.8 \pm 8.9$ \\
        \midrule
        across $\lambda$ & $\lambda = 1$ & $\bm{1.39 \pm 0.05}$ & $\bm{81.2 \pm 1.8}$ & \textit{1.58 $\pm$ 0.07} & $\bm{165.6 \pm 5.7}$ & $\bm{1.40 \pm 0.07}$ & $\bm{157.9 \pm 4.4}$ \\
         & $\lambda = 1/4$ & $1.61 \pm 0.05$ & $104.8 \pm 3.1$ & $\bm{1.42 \pm 0.07}$ & $236.0 \pm 7.7$ & \textit{1.60 $\pm$ 0.07} & $227.6 \pm 4.6$ \\
\bottomrule
    \end{tabular}
    \label{tab:rank-lambda}
\end{table}

\begin{table}[!ht]
    \caption{Average integral rank and mean simulation calls per design of MLOE at each weight $\lambda$, on the three seeds of the selection runs of the SST tasks, mean $\pm$ standard error. All four weights are taken at the same warm-start training regime, mixing with $\rho = 0.9$ on the SST angle task and no mixing ($\rho = 1$) on the SST layout task. Lower is better; 1 is the best attainable and 4 the worst. The lowest values are marked in bold, the values within two combined standard errors~---~in italics.}
    \centering
    \begin{tabular}{@{}lcccc@{}}
        \toprule
        & \multicolumn{2}{c}{SST angle} & \multicolumn{2}{c}{SST layout} \\
        \cmidrule(lr){2-3} \cmidrule(lr){4-5}
        & rank & calls per design, $10^3$ & rank & calls per design, $10^3$ \\
        \midrule
        $\lambda = 1$ & $2.76 \pm 0.14$ & $\bm{182.3 \pm 7.4}$ & $\bm{2.16 \pm 0.44}$ & $\bm{178.9 \pm 17.2}$ \\
        $\lambda = 1/2$ & \textit{2.05 $\pm$ 0.21} & $210.2 \pm 0.5$ & \textit{2.57 $\pm$ 0.41} & \textit{226.0 $\pm$ 16.7} \\
        $\lambda = 1/4$ & $\bm{1.81 \pm 0.16}$ & $247.0 \pm 7.7$ & \textit{2.87 $\pm$ 0.66} & $246.4 \pm 12.8$ \\
        $\lambda = 1/8$ & $3.38 \pm 0.09$ & $283.8 \pm 17.9$ & \textit{2.40 $\pm$ 0.26} & $280.9 \pm 7.9$ \\
        \bottomrule
    \end{tabular}
    \label{tab:rank-lambda-scan}
\end{table}

\section{Ablation studies}\label{app:ablation}

The context is removed from MLOE: the meta-inference model receives the design parameters as inputs; however, it is trained on the samples of the current design only. In this setting, the transfer between designs is carried by the initial weights alone, and optimality across the optimization trajectory is not enforced.

We conduct this study on the SHiP tasks. The runs use the ten PRNG seeds of the test runs to allow for a direct comparison with the main experiments. The studies were performed under the warm-start training regimes of the continual strategy selected for the main experiments: mixing with $\rho = 0.9$ on the SST angle task, no mixing ($\rho = 1$) on the SST layout task.

\begin{figure}[p]
    \centering
    \includegraphics[width=\textwidth]{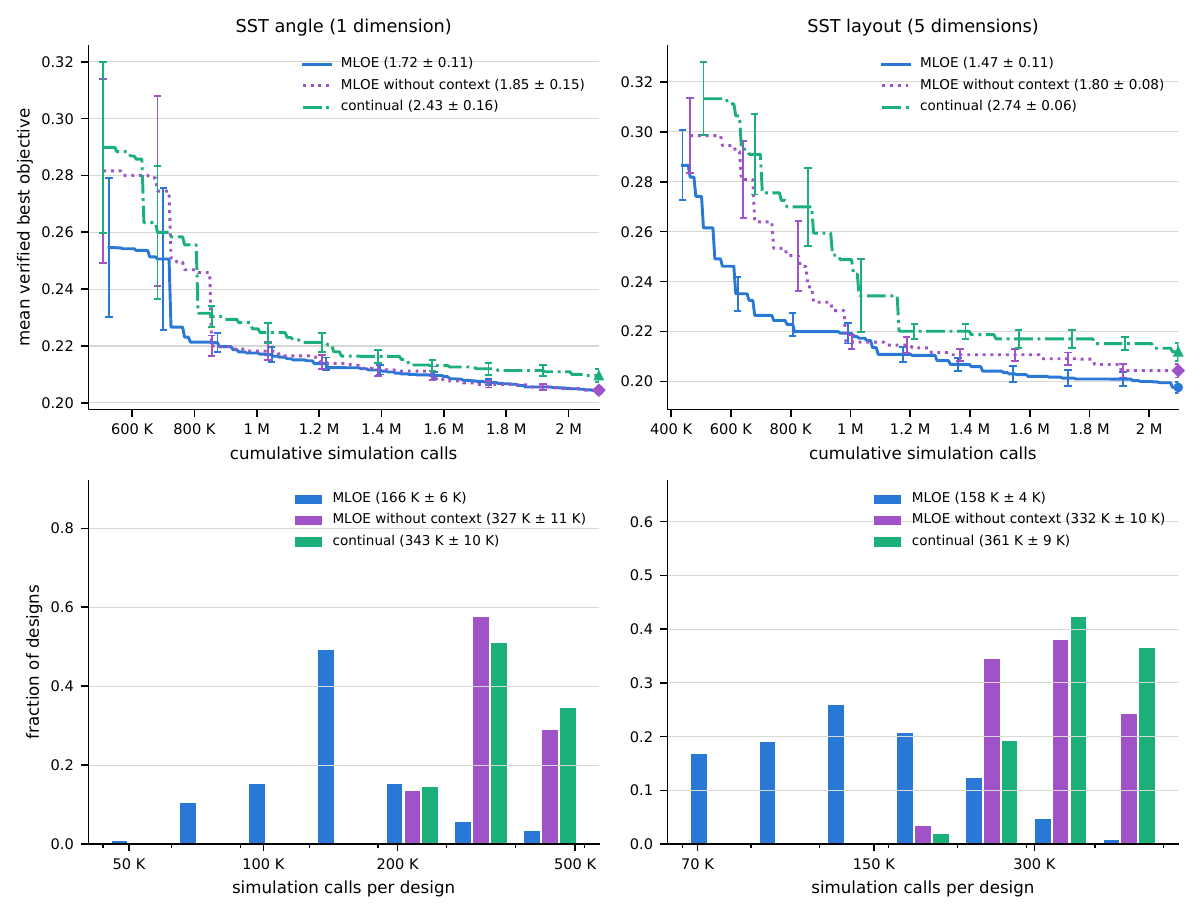}
    \caption{MLOE without the context, against MLOE and the continual strategy, on the SST angle (left) and SST layout (right) tasks. Top: verified best objective against cumulative simulation calls; average integral ranks are indicated in parentheses. Bottom: simulation calls per design; mean sample consumption per design is indicated in parentheses. Ten seeds.}
    \label{fig:ablation-no-context}
\end{figure}

Table~\ref{tab:rank-no-context} shows that removing the context leaves the convergence of MLOE intact, while the sample consumption per design rises towards that of the continual strategy. Figure~\ref{fig:ablation-no-context} shows that the rank comparison is dominated by the tails of the convergence curves where all strategies become statistically indistinguishable: MLOE with the full context converges faster in the middle of the trajectory, and MLOE without the context lies roughly between MLOE with the full context and the continual strategy, which suggests that the implicit transfer through the initial weights has a non-negligible effect, in line with the literature on catastrophic forgetting in continual learning~\cite{goodfellow2013empirical, ramasesh2021anatomy, mirzadeh2020understanding, evron2022catastrophic, lin2023theory}.

\begin{table}[!ht]
    \caption{Average integral rank and mean simulation calls per design of MLOE, MLOE without the context and the continual strategy, on the verified objective values over ten seeds, mean $\pm$ standard error. Lower is better; 1 is the best attainable and 3 the worst. The lowest values are marked in bold, the values within two combined standard errors~---~in italics.}
    \centering
    \begin{tabular}{@{}lcccc@{}}
        \toprule
        & \multicolumn{2}{c}{SST angle} & \multicolumn{2}{c}{SST layout} \\
        \cmidrule(lr){2-3} \cmidrule(lr){4-5}
        & rank & calls per design, $10^3$ & rank & calls per design, $10^3$ \\
        \midrule
        MLOE & $\bm{1.72 \pm 0.11}$ & $\bm{165.6 \pm 5.7}$ & $\bm{1.47 \pm 0.11}$ & $\bm{157.9 \pm 4.4}$ \\
        MLOE without context & \textit{1.85 $\pm$ 0.15} & $327.2 \pm 11.5$ & $1.80 \pm 0.08$ & $331.9 \pm 9.8$ \\
        continual & $2.43 \pm 0.16$ & $342.6 \pm 9.8$ & $2.74 \pm 0.06$ & $360.8 \pm 8.9$ \\
\bottomrule
    \end{tabular}
    \label{tab:rank-no-context}
\end{table}

\section{The source of acceleration}\label{app:source}
\begin{figure}[p]
    \centering
    \includegraphics[width=\textwidth]{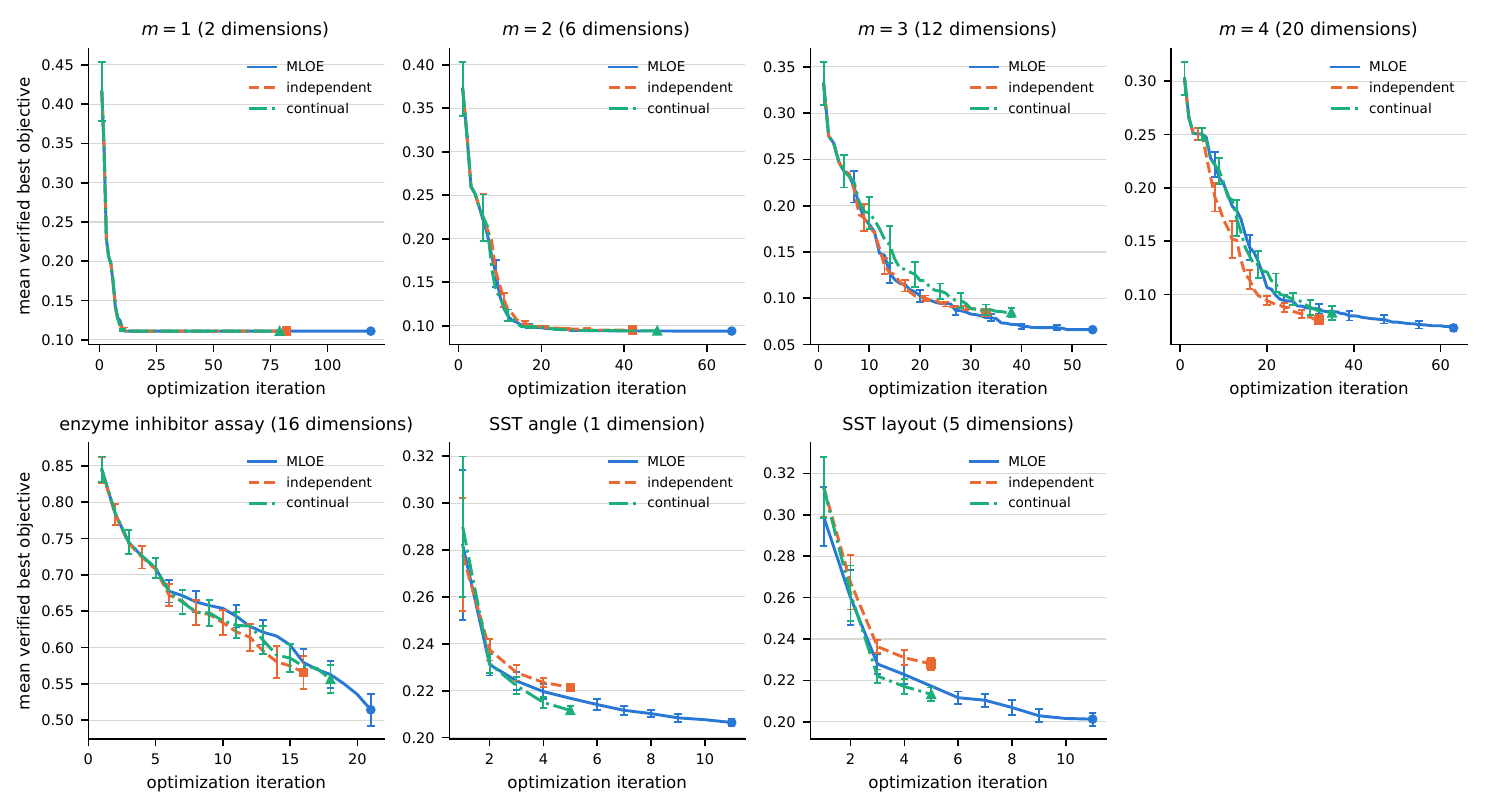}
    \caption{Mean verified best objective against the optimization iteration. Top row: probes for linear model, one panel per dimensionality $m$, ten seeds per dimensionality. Bottom row: enzyme inhibitor assay, twenty seeds; SST angle and SST layout, ten seeds per task. A curve is drawn while every seed has a design at that iteration.}
    \label{fig:iteration}
\end{figure}

This section examines the source of the accelerated convergence under MLOE. Figure~\ref{fig:iteration} shows the convergence curves as functions of the optimization iteration. In most cases, the curves of the strategies are indistinguishable, which indicates that the advantage of MLOE stems primarily from its lower sample consumption and, thus, from the larger number of iterations afforded by the budget. The only exception is the independent strategy on the SST tasks, which converges more slowly; the separation, however, lies within three standard errors. We hypothesize that the independent re-initialization increases the noise of the objective estimates~\cite{bouthillier2021accounting} and, thus, slightly degrades the Gaussian process surrogate, which often results in slower convergence~\cite{srinivas2012information}.

\begin{table}[ht]
    \caption{Cost of a single optimization step on the SST tasks using simplified simulations compared to estimated cost of generating the same number of samples using high-fidelity simulation.}
    \centering
    \begin{tabular}{@{}llccc@{}}
        \toprule
        & & & simplified & high-fidelity \\
        \cmidrule(lr){4-4} \cmidrule(lr){5-5}
        Task & Strategy & calls per design, $10^3$ & time per design, h & sampling, core-hours \\
        \midrule
        SST angle & MLOE & $166 \pm 6$ & $0.200 \pm 0.007$ & 92--138 \\
        & independent & $318 \pm 7$ & $0.337 \pm 0.011$ & 176--265 \\
        & continual & $343 \pm 10$ & $0.317 \pm 0.014$ & 190--286 \\
        \midrule
        SST layout & MLOE & $158 \pm 4$ & $0.189 \pm 0.007$ & 88--132 \\
        & independent & $324 \pm 6$ & $0.323 \pm 0.010$ & 180--270 \\
        & continual & $361 \pm 9$ & $0.323 \pm 0.009$ & 200--301 \\
        \bottomrule
    \end{tabular}
    \label{tab:timing}
\end{table}

\section{Wall-clock timing}\label{app:timing}

Throughout this work, the convergence of the strategies is computed against the cumulative number of simulation calls, assuming that the sampling of training data dominates the overall cost of an optimization run \emph{in practice}. To provide a rigorous comparison, we use simplified simulations that are much less demanding than the detailed, physically faithful versions. In this section, we estimate the expected computational time on the high-fidelity simulation, FairShip~\cite{fairship}, used by the SHiP collaboration~\cite{ahdida2022ship}.

Table~\ref{tab:timing} lists the cost of a single design on the SST tasks. With the simplified simulation a design takes approximately 10--20 minutes. The high-fidelity simulation transports the particles through the full SHiP geometry with Geant4 at approximately 2--3~s per event on a single CPU core (the rate depends on the hardware; the figure was measured for an AMD EPYC 7543), which would make the sampling time several orders of magnitude larger than those of acquisition function optimization and inference model training combined; therefore, the assumption largely holds in practical settings.

\clearpage


\clearpage
\clearpage

\clearpage

\bibliographystyle{unsrt}
\bibliography{main}